\documentclass[journal]{IEEEtran}
\usepackage{amsmath,amsfonts}
\usepackage{algorithmic}
\usepackage{algorithm}
\usepackage{array}
\usepackage[caption=false,font=normalsize,labelfont=sf,textfont=sf]{subfig}
\usepackage{textcomp}
\usepackage{stfloats}
\usepackage{url}
\usepackage{verbatim}
\usepackage{graphicx}
\usepackage{cite}
\usepackage{orcidlink}

\usepackage{makecell, tabularx}
\usepackage{tabularx}
\usepackage{amsmath}
\usepackage{amsfonts}
\usepackage{tikz}
\usepackage{multirow}
\usepackage{xcolor}
\usepackage[export]{adjustbox}
\usepackage{hyperref}

\begin{document}

\title{Beyond Subspace Isolation: \\ Many-to-Many Transformer for Light Field Image Super-resolution}

\author{Zeke Zexi Hu\orcidlink{0000-0003-4947-4832}, Xiaoming Chen\orcidlink{0000-0002-7503-3021},~\IEEEmembership{Member, IEEE}, Vera Yuk Ying Chung\orcidlink{0000-0002-3158-9650},~\IEEEmembership{Member, IEEE}, \\ Yiran Shen\orcidlink{0000-0003-1385-1480},~\IEEEmembership{Senior Member, IEEE}
\thanks{This work was supported in part by Beijing Natural Science Foundation (No. 4222003) and National Natural Science Foundation of China (No. 62177001). (Corresponding author: Xiaoming Chen.)}
\thanks{Zeke Zexi Hu and Vera Yuk Ying Chung are with the School of Computer Science, University of Sydney, Darlington, NSW 2008, Australia (e-mail: zexi.hu@sydney.edu.au; vera.chung@sydney.edu.au).}
\thanks{Xiaoming Chen is with the School of Computer Science and Engineering, Beijing Technology and Business University, Beijing 102488, China (e-mail: xiaoming.chen@btbu.edu.cn).}
\thanks{Yiran Shen is with the School of Software, Shandong University, Jinan, 250100, China (e-mail: yiran.shen@sdu.edu.cn).}
}

\markboth{Journal of \LaTeX\ Class Files,~Vol.~14, No.~8, August~2021}%
{Shell \MakeLowercase{\textit{et al.}}: A Sample Article Using IEEEtran.cls for IEEE Journals}


\newif\ifhl

\ifhl
    \newcommand{\hl}[1]{\textcolor{blue}{#1}}  
    \newenvironment{highlight}{\color{blue}}{}
\else
    \newcommand{\hl}[1]{#1}
    \newenvironment{highlight}{}{}
\fi

\maketitle

\begin{abstract}
    The effective extraction of spatial-angular features plays a crucial role in light field image super-resolution (LFSR) tasks, and the introduction of convolution and Transformers leads to significant improvement in this area. Nevertheless, due to the large 4D data volume of light field images, many existing methods opted to decompose the data into a number of lower-dimensional subspaces and perform Transformers in each sub-space individually. As a side effect, these methods inadvertently restrict the self-attention mechanisms to a One-to-One scheme accessing only a limited subset of LF data, explicitly preventing comprehensive optimization on all spatial and angular cues. In this paper, we identify this limitation as subspace isolation and introduce a novel Many-to-Many Transformer (M2MT) to address it. M2MT aggregates angular information in the spatial subspace before performing the self-attention mechanism. It enables complete access to all information across all sub-aperture images (SAIs) in a light field image. Consequently, M2MT is enabled to comprehensively capture long-range correlation dependencies. With M2MT as the foundational component, we develop a simple yet effective M2MT network for LFSR. Our experimental results demonstrate that M2MT achieves state-of-the-art performance across various public datasets, and it offers a favorable balance between model performance and efficiency, yielding higher-quality LFSR results with substantially lower demand for memory and computation. We further conduct in-depth analysis using local attribution maps (LAM) to obtain visual interpretability, and the results validate that M2MT is empowered with a truly non-local context in both spatial and angular subspaces to mitigate subspace isolation and acquire effective spatial-angular representation.
\end{abstract}


\section{Introduction} \label{section:Introduction}

Light field (LF) images, unlike regular images captured by monocular cameras, provide richer information by capturing light rays from multiple angular directions in a single shot. This unique characteristic has paved the way for substantial progress in various computer vision applications where conventional cameras have shown limited efficacy, \textit{e.g.,} material recognition \cite{wangLFRecognition_ECCV2016, luLFRecognition_2019}, depth estimation \cite{yucer2016efficient, heberUshapeICCV2017, wangOcclusionawareDepthEstimation2015, chaoSubFocal_TCI2023, ding_TCI2023}, salient object detection under complex scenarios \cite{shengLFSaliency_ICASSP2016, zhangLFNet_TIP2020, chen_TMM2023}, microscopy \cite{verinaz_TCI2022, verinaz_TCI2023, levoy2006light}, and anti-spoof face recognition \cite{raghavendraLFFace_TIP2015, jiLFHOG_ICIP2016}. By simultaneously capturing multiple sub-aperture images (SAIs, or views), LF technology enables a rich and interactive viewing experience. Users can freely explore and interact with the virtual environments, changing perspectives and moving within them. Therefore, LF technology has become a cornerstone of VR applications, enhancing user engagement and immersion.

\newcommand{\imageWithGrid}[3]{%
  \begin{tikzpicture}
    \node[anchor=south west,inner sep=0] (image) at (0,0) {\includegraphics[width=#2, height=#3]{#1}};
    \begin{scope}[x={(image.south east)},y={(image.north west)}]
        \foreach \i in {1,...,4} {
            \draw[lightgray,thin] (\i/5,0) -- (\i/5,1); 
            \draw[lightgray,thin] (0,\i/5) -- (1,\i/5); 
        }
        \draw[black] (0,0) rectangle (1,1);
    \end{scope}
  \end{tikzpicture}%
}

\begin{figure}[t!]
\centering

\tabcolsep=0.04cm
\renewcommand{\arraystretch}{0.8}
\begin{tabular}{ccccc}
    \raisebox{0.4\height}{
        \resizebox{0.05\textwidth}{!}{
        \begin{tikzpicture}
            \foreach \x in {0,1,2,3,4} {
                \foreach \y in {0,1,2,3,4} {
                    \draw[black, thin] (\x,\y) rectangle (\x+1,\y+1);
                }
            }
            \draw[black] (0,0) rectangle (5,5);
            \fill[red] (2,2) rectangle (3,3);
        \end{tikzpicture}}
    } &
    \includegraphics[width=0.09\textwidth]{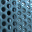} &
    \includegraphics[width=0.09\textwidth]{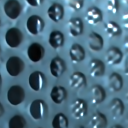} &
    \includegraphics[width=0.09\textwidth]{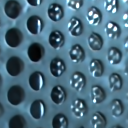} &
    \includegraphics[width=0.09\textwidth]{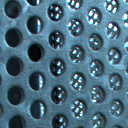} \\
    \footnotesize{\makecell{SAI\\Location}} & \footnotesize LR & \footnotesize EPIT & \footnotesize M2MT-Net & \footnotesize HR
\end{tabular} \\

(a) SAI location and patch images. \\
\hspace{0pt}

\tabcolsep=0.1cm
\renewcommand{\arraystretch}{0.9}
\begin{tabular}{cc}
    \imageWithGrid{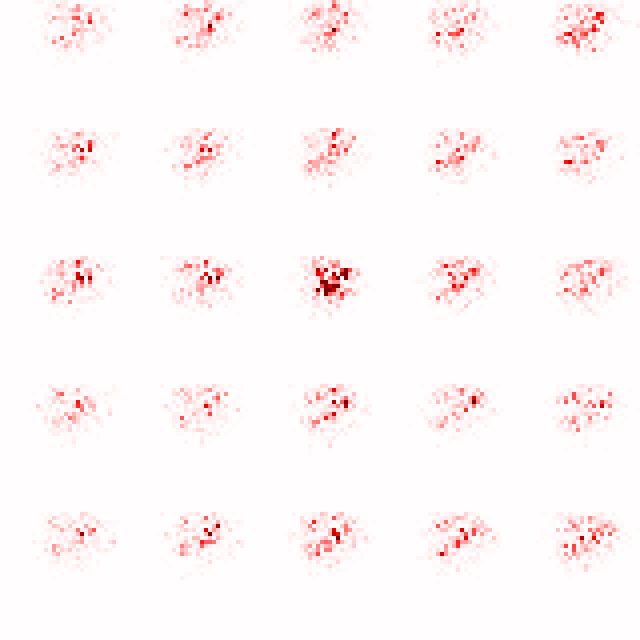}{0.18\textwidth}{0.18\textwidth} &
    \imageWithGrid{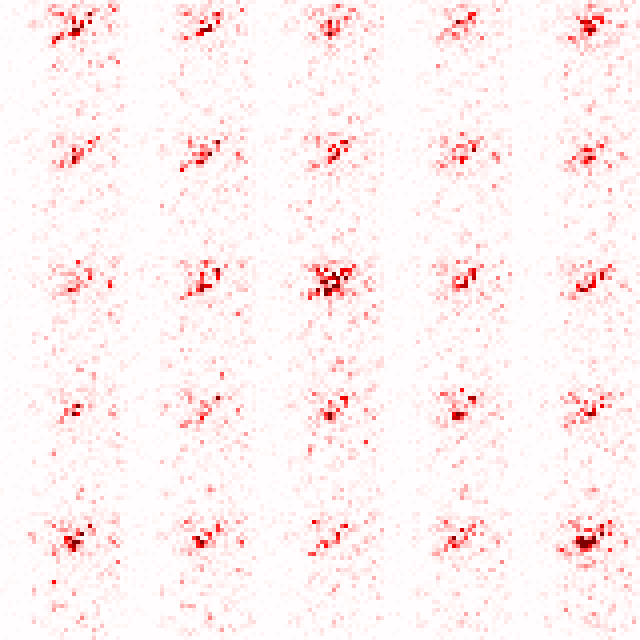}{0.18\textwidth}{0.18\textwidth} \\
    EPIT & M2MT-Net \\
    DI: 8.3895 & DI: 25.1975 \\
\end{tabular} \\
\vspace{8pt}
(b) Local attribution maps of SAIs. Diffusion Index (DI) quantifies the extent of influential pixels. \\
\hspace{0pt}
\caption{Super-resolution results and local attribute maps (LAM) of the proposed M2MT-Net against EPIT on the \textit{Perforated\_metal\_3} sample.}

\label{fig:First}
\end{figure}

Capturing LF images necessitated self-built dense camera arrays \cite{wilburn2004high, wilburn2005high}, which were prohibitively expensive and not ready for mainstream use. However, advancements in sophisticated LF cameras like Raytrix \cite{Raytrix}, Lytro Illum \cite{Lytro}, and Google's Light Field VR Camera \cite{GoogleLF} have gradually democratized LF imaging, making it accessible for both commercial and industrial applications. Despite this progress, LF cameras have long faced challenges in striking a balance of angular and spatial resolutions due to inherent limitations in sensor capabilities, often leading to lower spatial resolutions compared to traditional cameras.

Researchers have developed a number of possible solutions, and they generally fall into two categories: light field image super-resolution (LFSR) \cite{yeungSAS_LFSR2019,wangDistgSSR_TIP2022, congLFDET_TMM2024} \hl{and light field view synthesis (also known as light field reconstruction or light field angular super-resolution) \cite{wuSAAN_TIP2021, liu_TMM2022, yeungSAS_ECCV2018, chengLFSSRSAV_TCI2022, huSADenseNet_TIM2021, wuShearedEPI_TIP2019}}. LFSR aims to enhance the spatial resolution of all SAIs, while light field view synthesis focuses on synthesizing additional SAIs to enhance the angular resolution of a light field image. \hl{Additionally, some works \cite{yoon2015LFCNN, koRemixing_TIP2021, wang2023mfsrnet} have developed methods for joint super-resolution that enhance both spatial and angular resolutions simultaneously.} In this paper, we primarily concentrate on LFSR.

\begin{figure*}[t!]
    \definecolor{myred}{HTML}{FF0080}
    \definecolor{myblue}{HTML}{007FFF}
    \centering

    \tabcolsep=0.10cm
    \renewcommand{\arraystretch}{0.1}
    \setlength{\medmuskip}{-1mu}
    \begin{tabular}{cc}
        \raisebox{0.025\height}{
        \includegraphics[width=0.250\linewidth]{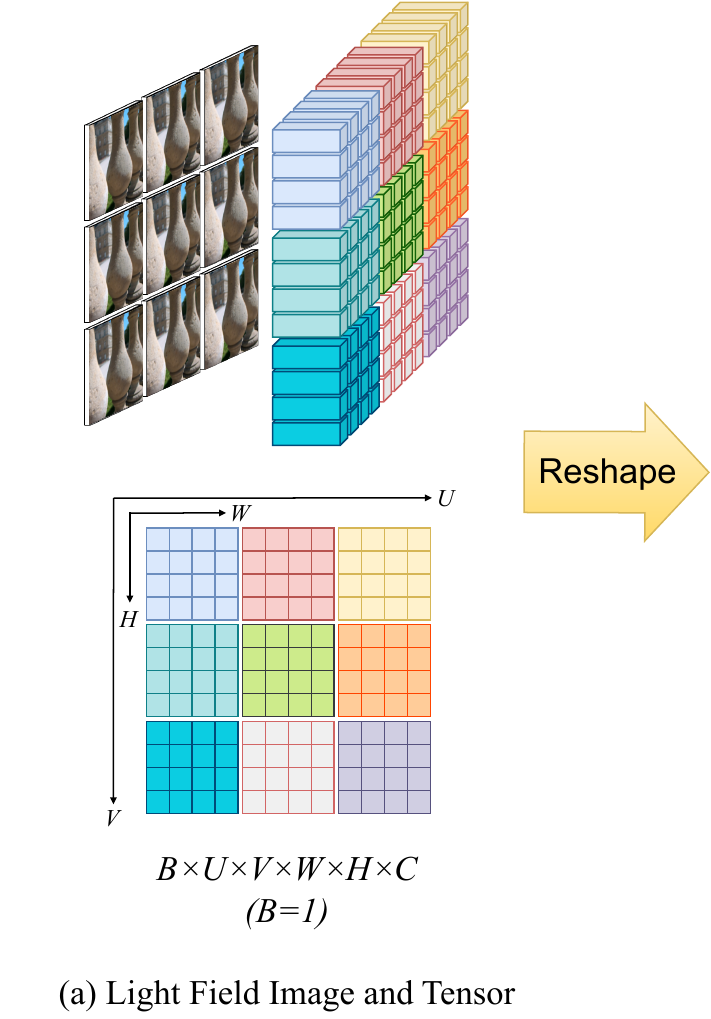}
        }
        &
        \includegraphics[width=0.660\linewidth]{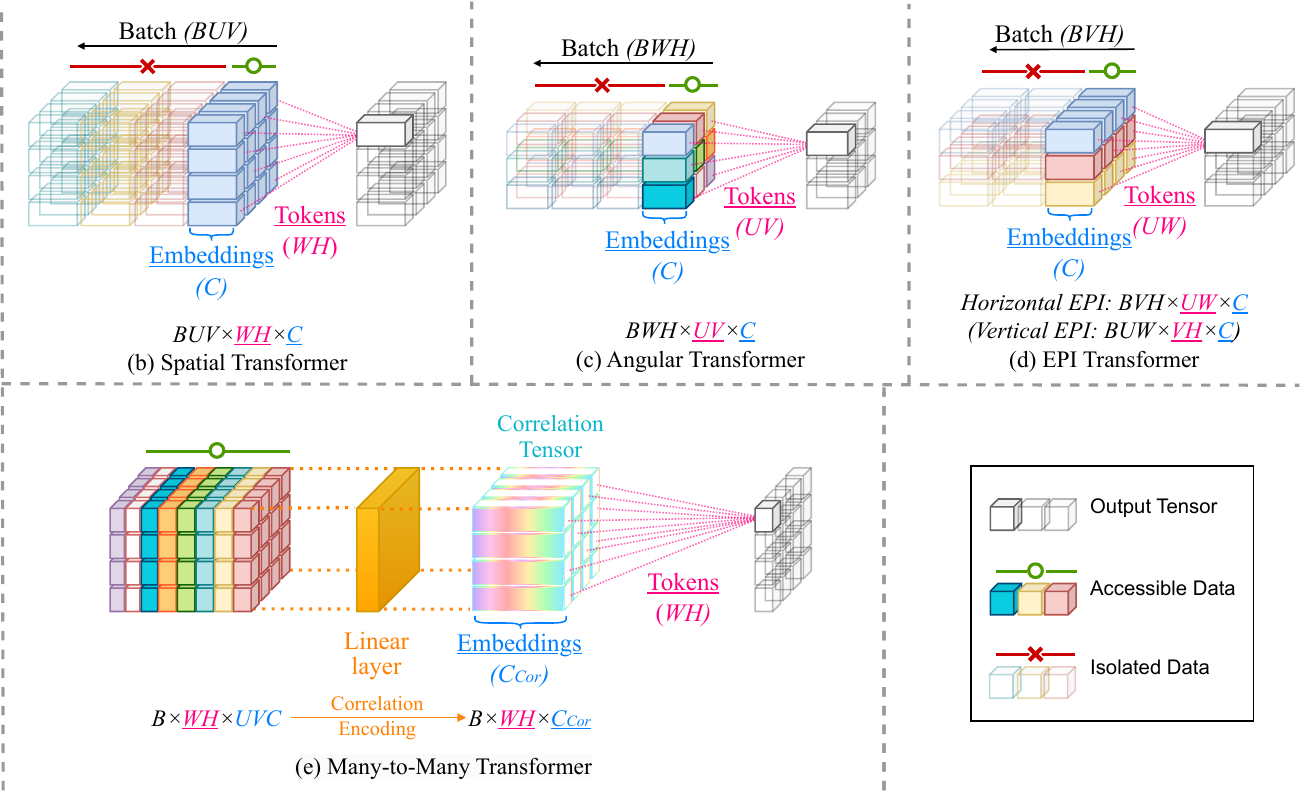}
    \end{tabular}
    \caption{Illustration of accessible data in LF tensors used by existing LF Transformers under the One-to-One scheme and our proposed Many-to-Many Transformer. For the LF tensors, each color represents a SAI.}
    \label{fig:Cubes}
\end{figure*}

\noindent{\bf Motivations.} The advances in deep learning, particularly convolutional neural networks (CNNs) \cite{dongSRCNN_ECCV2014, zhangRCAN_ECCV2018} and Vision Transformers (ViT) \cite{ViT, liuSwin_ICCV2021, luESRT_CVPR2022}, have led to significant improvement in LFSR than traditional methods \cite{wanner2014variational}. Among them, most methods have opted to process 4D LF images in their 2D subspaces, such as spatial, angular \cite{wangDPT_AAAI2022, liangLFT_SPL2022, congLFDET_TMM2024, wangDistgSSR_TIP2022}, or Epipolar Image (EPI) \cite{liangEPIT_ICCV2023, wangDistgSSR_TIP2022} subspaces. However, these methods predominantly suffer from subspace isolation, a defect causing sub-optimal performance.

Specifically, when adapting 2D operations to the 4D LF data, existing methods have to compromise their complete access to the LF information. This is primarily due to training networks directly on voluminous 4D LF data, e.g., 4D convolutions \cite{yeungSAS_LFSR2019}, demands a relatively large number of weights, which is prone to optimization difficulties and heavy computation. As a workaround, most previous methods decompose the 4D data into 2D subspaces such as spatial and angular subspaces, or EPI subspaces. In implementation, one typical practice is to temporarily reshape a 4D tensor to expose two operative dimensions while merging the other two dimensions with the batch dimension. This decomposition enables 2D operations to perform on 4D LF tensors, and in network training, the optimization is conducted on the whole tensor. However, a significant limitation arises during inference. When inferring the value at a specific location, access to the two merged dimensions is confined to only one location at a time rather than spanning the entire dimensionality. As a result, even with non-local Transformers, the effective receptive field is virtually restricted to a local context within the operative subspaces, leading to a local One-to-One scheme.

\hl{For instance, considering the scenario depicted in Fig. \ref{fig:Cubes}(a), a single 4D LF tensor within a batch $B$ is defined as $I(u, v, x, y) \in \mathbb{R}^{U \times V \times W \times H \times C}$, where $U$ and $V$ denote the two angular dimensions, $W$ and $H$ denote the two spatial dimensions, and $C$ denotes the channel dimension.} Here, $(u, v, x, y)$ denotes a pixel's spatial location $(x, y)$ and angular location (or SAI) $(u, v)$. By merging the angular subspace $(U \times V)$ into the batch dimension $B$, a 2D spatial Transformer in Fig. \ref{fig:Cubes}(b) is enabled to operate on the flattened spatial subspace $(W \times H)$ as tokens across SAIs. However, this merging operation virtually isolates the network's forward propagation within only a location in the merged angular subspace $BUV$. The accessible data in the batch dimension is depicted by the opaque block, while the isolated data is transparent. Consequently, the network's receptive field is restricted to only one SAI at a time during inference. This procedural constraint can be formally expressed as
\begin{equation} \label{eq:before}
\begin{split}
 I_{2}(u, v, x, y) = \hl{F_{O2O}} \cdot \{I_{1}(\bar{u}, \bar{v}, \bar{x}, \bar{y})\}_{(\bar{u}, \bar{v}) = (u, v), (\bar{x}, \bar{y}) \in \mathbb{R}^{W \times H}}
\end{split}
\end{equation}
where \hl{$F_{O2O}$ represents a One-to-One operation}, which can be either convolution or Transformers, and $I_{1}$ and $I_{2}$ are the input and output LF tensors of the operation. Under this scheme, to obtain a complete LF tensor, Equation \ref{eq:before} must be repeated $U \times V$ times mapping from a SAI in $I_{1}$ at a single angular location $(\bar{u}, \bar{v})$ to a SAI in $I_{2}$ at the same isolated angular location $(u, v)$ in the output. However, the ideal processing would instead use all SAIs to inform the calculation loosening the constraint $(\bar{u}, \bar{v}) = (u, v)$, \hl{resulting in a Many-to-Many operation $F_{M2M}$}:
\begin{equation} \label{eq:after}
\begin{split}
 I_{2}(u, v, x, y) = \hl{F_{M2M}} \cdot \{I_{1}(\bar{u}, \bar{v}, \bar{x}, \bar{y})\}_{(\bar{u}, \bar{v}, \bar{x}, \bar{y}) \in \mathbb{R}^{U \times V \times W \times H}}.
\end{split}
\end{equation}

\hl{Subspace isolation is not unique to spatial Transformers and extends to other forms of data decomposition under the One-to-One scheme. For example, an angular Transformer is limited to accessing only one pixel across SAIs, as depicted in Fig. \ref{fig:Cubes}(c). Similarly, an EPI Transformer can only process a two-dimensional slice within the EPI subspace at one step. Specifically, it can handle a horizontal EPI $\mathbb{R} \in R^{U \times W}$ or a vertical EPI $\mathbb{R} \in R^{V \times H}$ as illustrated in Fig. \ref{fig:Cubes}(d), while the remaining slices stay isolated in the batch dimension.} These constraints, inherent in the local One-to-One operational scheme, significantly impede the ability of existing models to fully exploit the spatial and angular cues available in LF data, resulting in an incomplete spatial-angular representation.

\noindent{\bf Contributions.}
To address this issue, in this paper, we propose the novel Many-to-Many Transformer (M2MT), a new scheme to achieve the goal of comprehensive data integration outlined in Equation \ref{eq:after} and alleviate the isolation. The M2MT method begins by constructing a correlation tensor in the angular subspace. It then applies a self-attention mechanism to model long-range dependencies within the spatial subspace. This innovative approach allows the M2MT to access all the spatial and angular cues present in an LF image during each step of data propagation, thereby facilitating the creation of a comprehensive spatial-angular representation with a truly non-local context.

With M2MT as a foundational component, we present a simple yet effective network, M2MT-Net, incorporating M2MT in the spatial subspace and vanilla Transformers in the angular subspace. Through extensive experimental evaluation, we showcase M2MT-Net's outstanding performance and an excellent performance-efficency balance, establishing it as a new state-of-the-art for LFSR.

Furthering the research, we delve into a series of studies to discover the mechanisms behind its success. Particularly, by leveraging the technique of local attribution maps (LAM) \cite{guLAM_CVPR2021}, which visualize influential pixels, to gain interpretability of neural networks. Fig. \ref{fig:First} reveals that M2MT-Net utilizes more pixels across broader SAIs than the current state-of-the-art methods like EPIT \cite{liangEPIT_ICCV2023}. This observation substantiates the efficacy of M2MT-Net, which mitigates the limitation of subspace isolation, simultaneously preserving more high-frequency cues in the spatial subspace and establishing a richer and non-local interplay of SAI dependencies in the angular subspace.

We also conducted an analysis using light field depth estimation to validate the angular consistency in the reconstructed results. The results demonstrate that M2MT-Net's depth maps are sharper and more integrated, suggesting that, besides reconstructing more visual details, M2MT-Net effectively preserves the parallax structure across SAIs, enriching the realism of the resulting LF images.

The contributions of this paper can be summarized as follows:
\hl{
\begin{enumerate}
    \item We propose the Many-to-Many Transformer (M2MT), a novel approach integrating spatial and angular information in light field images. By constructing a correlation tensor in the angular subspace and applying a self-attention mechanism in the spatial subspace, M2MT addresses the subspace isolation prevalent in the previous methods by its truly non-local context.
    \item We introduce M2MT-Net, which incorporates M2MT in the spatial subspace and vanilla Transformers in the angular subspace. Extensive experiments show that M2MT-Net sets a new state-of-the-art for LFSR in terms of performance. Furthermore, M2MT-Net strikes a compelling balance between model performance and efficiency, producing higher-quality LFSR results with substantially lower memory and computation requirements.
    \item We provide insights into M2MT-Net's effectiveness. The analysis of local attribution maps (LAM) is conducted to visualize influential pixels, showing that M2MT-Net utilizes more pixels across a broader range of sub-aperture images (SAIs) compared to existing methods. Additionally, our analysis of light field depth estimation reveals that M2MT-Net produces sharper and more integrated depth maps, which suggests that it preserves the parallax structure of LF images, enhancing its realism with better angular consistency.
\end{enumerate}
}
\section{Related Work}
\subsection{Single Image Super-resolution}
Single Image Super-resolution (SISR) is a classic low-level computer vision task aiming to reconstruct a high-resolution image (HR) from the low-resolution (LR) counterpart. Dong et al. \cite{dongSRCNN_ECCV2014} pioneered the introduction of CNN to this task, setting a new standard that outperformed previous SISR methods. This innovation marked the inception of a trend in the realm towards the widespread integration of deep neural networks. Subsequent achievements include VDSR \cite{kimVDSR_CVPR2016}, which leverages the residual connection to improve the data flow in a deep neural network; RDN \cite{zhangRDN_CVPR2018}, similarly improving the data flow via densely connected networks; and RCAN \cite{zhangRCAN_ECCV2018}, incorporating a residual-in-residual structure to further amplify the benefits of residual connections. Some works explored to utilize information in other domains for SISR, such as spectral information \cite{esmaeilzehi_TCI2021} and text-to-image models \cite{wuSeeSR_arXiv2023}.

Other contributions, such as SRGAN \cite{ledigSRGAN_CVPR2017} and EnhanceNet \cite{SajjadiEnhanceNet_ICCV2017}, emphasized the generation of visually appealing details by training networks using feature-based loss functions or adversarial learning. More recently, drawing inspiration from the success of Vision Transformer (ViT) \cite{ViT} in high-level vision tasks, Transformer-based SISR methods have further enhanced SISR by leveraging the self-attention mechanism. IPT \cite{chenIPT_CVPR2021} introduced image processing Transformers pre-trained across image processing tasks to benefit from datasets for not only SISR but also image denoizing and image restoration. SwinIR \cite{liangSwinIR_ICCV2021} introduced the Swin Transformer \cite{liuSwin_ICCV2021}, a shifted window scheme, to a series of low-level vision tasks. HAT \cite{chenHAT_CVPR2023} proposed a hybrid attention component that combines channel attention convolution and window-based Transformers to enable the capability of global statistics and local fitting. Despite their success, Transformers are inherently accompanied by a quadratic growth in computational complexity relative to the input image size, which remains a challenge to their applicability in SISR. In response to this challenge, studies such as SRFormer \cite{zhouSRFormer_ICCV2023} and ELAN \cite{zhangELAN_ECCV2022} have emerged, aiming to alleviate the computational burden. SRFormer \cite{zhouSRFormer_ICCV2023} achieved this through permuted self-attention, while ELAN \cite{zhangELAN_ECCV2022} employed a long-range attention mechanism.

Different from the sole focus of SISR on enhancing visual details destroyed in downsampling, the LFSR task aims not only to restore these details but also to maintain and improve the parallax structure across SAIs, enriching the realism of the resulting LF images.

\subsection{Light Field Image Super-resolution}
Processing 4D LF data presents significant challenges in developing neural networks. The application of 4D convolutions is a straightforward solution but results in computationally heavy models, making both training and inference difficult.

To alleviate this drawback, \hl{Farrugia et al. \cite{farrugiaLFSR_TPAMI2019} proposed a framework incorporating optical flow and a deep CNN to reduce the angular subspace to construct a compact representation preserving angular consistency and subsequently restore the whole LF image.} Wang et al. \cite{wangLFRecognition_ECCV2016} introduced an interleaved filter as an approximation for light field material recognition. \hl{The filter decomposes a 4D convolution into a spatial convolution and an angular convolution.} They proved that comparable performance can be achieved by interleaving these two distinct convolutions.

This decomposition scheme was adopted by Yoon et al. in LFCNN for LFSR \cite{yoon2017LFCNN}. LFCNN consists of a spatial sub-network for SAI processing and an angular sub-network composed of three branches to capture LF correlation in three different geometric directions. Yeung et al. \cite{yeungSAS_LFSR2019} proposed a deep neural network consisting of a series of spatial-angular separable (SAS) convolution, akin to interleaved filters but trained in an end-to-end manner. Jin et al. \cite{jinLFSSRATO_2020} proposed an all-to-one framework where each SAI is individually super-resolved using the other SAIs. A structure-aware loss is incorporated to preserve LF images' inherent parallax structure. Wang et al. \cite{wangLfInterNet_ECCV2020} introduced a network to extract spatial and angular features in separate branches and iteratively fuse them. Liu et al. \cite{liuLFIINet_TMM2021} proposed a pyramid network with dilated convolutions to expand receptive fields in both spatial and angular subspaces. Chen et al. \cite{chen_TMM2021} incorporated the frequency domain and semantic prior and proposed a network to super-resolve both spatial and angular resolutions. Sun et al. \cite{sun_TMM2022} proposed a network with disparity-exploited and non-disparity branches to learn a compact spatial-angular representation. 

Further advancing the scheme, \hl{Cheng et al. \cite{chengLFSSRSAV_TCI2022} proposed the concept of spatial-angular correlated convolution, extending the SAS scheme \cite{yeungSAS_LFSR2019} to the EPI subspaces and forming spatial-angular versatile convolution (LFSSR-SAV).} Hu et al.\cite{huDKNet_TIM2022} proposed the Decomposition Kernel Network, which generalizes the decomposition scheme to comprehensively cover the spatial, angular and EPI subspaces. Wang et al. \cite{wangDistgSSR_TIP2022} proposed a disentangling mechanism to aggregate and enhance features from these subspaces. \hl{Duong et al. \cite{duongHLFSR_TCI2023} combined the angular and spatial extractors with its proposed multi-orientation epipolar extractors to cover more aspects of LF images.}

Different from the previous methods, some works resort to non-deep-learning-based models \cite{rossi2018geometry, ghassab_TMM2019}. Some works explored plug-and-play strategies to boost the performance of existing methods, like the learning prior from single images \cite{wangBoosting_TCI2023} and the cut-and-blend data augmentation \cite{xiaoCutMIB_CVPR2023}.

In parallel to SISR, ViT has broadened the LFSR landscape. DPT \cite{wangDPT_AAAI2022} leveraged Transformers to learn image and gradient information among SAIs in horizontal and vertical sequences. LFT \cite{liangLFT_SPL2022} drew parallels with the earlier decomposition scheme but employed Transformers in place of separable convolutions. To enable spatial Transformers to model both local and non-local dependencies, the spatial features were locally unfolded into patches and subsequently processed through a linear layer into local embeddings before self-attention. Liang et al. proposed EPIT \cite{liangEPIT_ICCV2023} to further explore the use of Transformers in horizontal and vertical EPI subspaces. \hl{To enhance the capability of spatial and angular Transformers, Cong et al. proposed a sub-sampling spatial modeling strategy and a multi-scale angular modeling strategy in their LF-DET \cite{congLFDET_TMM2024}}.

Despite these advancements, a common limitation predominantly persists across most decomposition-based methods: subspace isolation, as elaborated in Section \ref*{section:Introduction}. This limitation motivates the derivation of our work in this paper.

\begin{figure*}[ht!]
    \centering
    \includegraphics[width=0.95\textwidth]{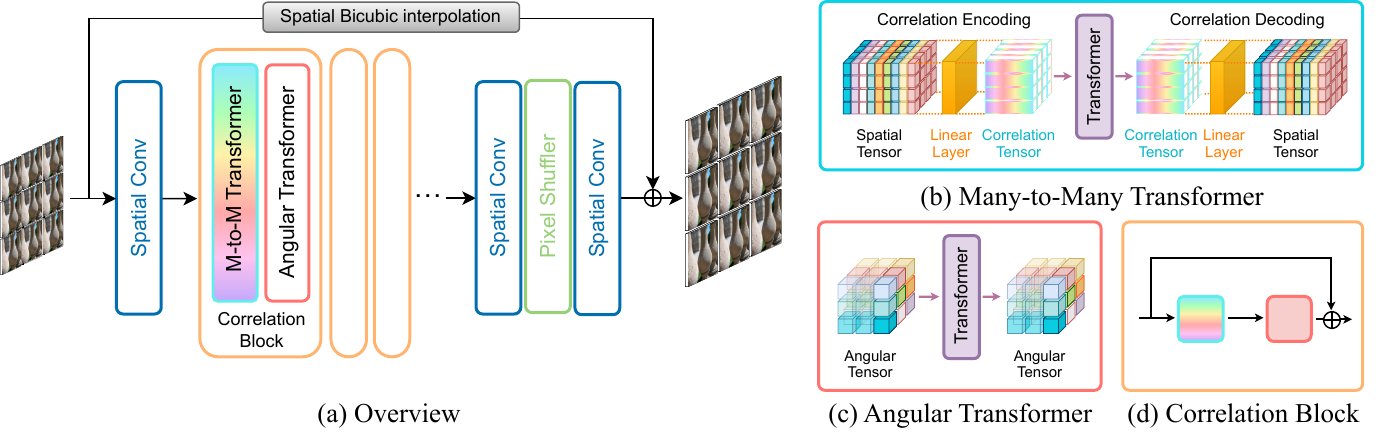}
    \vspace{-6pt}
    \caption{Illustration of M2MT-Net and its components. (a) depicts the overview of M2MT. (b) and (c) illustrate the details of a M2MT Transformer and an angular Transformer. These two components constitute a Correlation Block in (d). $\bigoplus$ represents the addition operation of a residual connection.} \label{fig:M2MT}
\end{figure*}

\section{Methodology}
\subsection{Problem Formulation} \label{section:Preliminary}
In formal terms, the procedure of LFSR is to enhance the spatial resolution from a low-resolution (LR) LF image $I_{LR}$ to a super-resolved (SR) LF image $I_{SR}$, which serves as an approximation to the corresponding high-resolution (HR) LF image $I_{HR}$. It can be denoted as
\begin{equation}\label{eq:obj}
\begin{split}
 I_{SR} = \mathcal{F}(I_{LR}), & \quad I_{LR}(u,v,x,y) \in \mathbb{R}^{U \times V \times W \times H \times C}, \\
    & \quad I_{SR}(u,v,x,y) \in \mathbb{R}^{U \times V \times rW \times rH \times C}
\end{split}
\end{equation}
\hl{where $\mathcal{F}(\cdot)$ is the super-resolution process,} $(U \times V)$ and $(W \times H)$ stand for the LR image's angular and spatial resolutions, respectively, $C$ denotes the channel dimension, $(u, v)$ indicates an angular location, $(x, y)$ indicates a spatial location, and $r$ represents the scale factor. 

A LF image or tensor, as shown in Fig. \ref{fig:Cubes}(a), can be reshaped into various forms to reveal distinctive subspaces. These encompass a spatial tensor $I_S$, revealing the spatial subspace $(W \times H)$ depicted in Fig. \ref{fig:Cubes}(b), an angular tensor $I_A$ in the angular subspace $(U \times V)$ depicted in Fig. \ref{fig:Cubes}(c), and EPI tensors $I_{EPI}$ which expose an EPI subspace, which consists of a spatial dimension and an angular dimension. Fig. \ref{fig:Cubes}(d) illustrates the tensor exposing $(U \times W)$ or $(V \times H)$, two typical EPI subspaces.

\subsection{Network Architecture}
The architecture of our proposed M2MT-Net is depicted in Fig. \ref{fig:M2MT}(a). It adopts a streamlined yet effective design comprising three phases. The first phase involves initial feature extraction, accomplished through $n_1$ spatial convolution layers. The crux of our architecture, the second phase, encompasses a sequence of $n_2$ correlation blocks. Each block incorporates two distinctive Transformers, namely a Many-to-Many Transformer (M2MT) and an angular Transformer, operating consecutively. A simplified visualization of a correlation block is given in Fig. \ref{fig:M2MT}(d). The last phase is pixel generation, which upsamples the extracted features by expanding the channel dimensions by $r^2$ times with a $1 \times 1$ convolution, followed by a pixel shuffler to increase the spatial resolution from $U \times V \times W \times H \times (r^2 C)$ to $U \times V \times rW \times rH \times C$, and lastly, a $3 \times 3$ convolution to squeeze the channels. Additionally, residual learning is enforced to allow the network to effectively capture residual information by learning from the differences between the HR and the bicubic-interpolated LR input. Also, within each correlation block, a residual skip connection is utilized to improve the information flow.

\subsection{Many-to-Many Transformer}
As the pivotal component, M2MT is proposed to mitigate the challenge posed by subspace isolation. Its objective is to holistically extract spatial-angular features with all spatial and angular cues from a LF image.

\begin{highlight}
A general Transformer \cite{vaswaniAttentionAllYou_NeurIPS2017} processes an input tensor $X \in \mathbb{R}^{B \times L \times D}$, where $B$ represents the batch dimension, $L$ is a sequence of tokens, and each token is a $D$-dimensional embedding. The Transformer's self-attention mechanism captures long-range dependencies by integrating information across all $L$ tokens globally.

To handle a 4D LF image $I \in \mathbb{R}^{U \times V \times W \times H \times C}$ using the spatial subspace as tokens, as illustrated in Fig. \ref{fig:Cubes}(b), conventional approaches \cite{congLFDET_TMM2024,liangLFT_SPL2022} merge the angular subspace with the batch dimension, resulting in a spatial tensor $I_{S} \in \mathbb{R}^{BUV \times WH \times C}$, where $WH$ serves as tokens and $C$ serves as embeddings ($L=WH$ and $D=C$). However, this method leads to subspace isolation, as discussed in Section \ref{section:Introduction}.

To address this issue, the proposed M2MT diverges from this conventional approach. A simplified illustration of M2MT is depicted in Fig. \ref{fig:M2MT}(b), and a detailed one in Fig. \ref{fig:Cubes}(e). Specifically, it initiates by merging the angular subspace with the channel dimensions, yielding a spatial tensor in a special form $I_{\tilde{S}} \in \mathbb{R}^{B \times WH \times UVC}$, which prepares the tensor for the following correlation encoding process. The correlation encoding process transforms $I_{\tilde{S}}$ into a correlation tensor $I_{Cor} \in \mathbb{R}^{B \times WH \times C_{Cor}}$:

\begin{equation} \label{eq:encoding1}
I_{Cor} = L_{encode}(I_{\tilde{S}})
\end{equation}
where $C_{Cor}$ denotes the number of channels of the correlation tensor, and the correlation encoding process $L_{encode}: \mathbb{R}^{B \times WH \times UVC} \mapsto \mathbb{R}^{B \times WH \times C_{Cor}}$ is implemented through a linear layer (or a fully connected layer) with a weight matrix $W_{encode} \in \mathbb{R}^{UVC \times C_{Cor}}$:
\begin{equation} \label{eq:encoding2}
 L_{encode}(X) = W_{encode} X.
\end{equation}

The resultant $I_{Cor}$ aggregates the angular correlation information at each spatial location into a compact feature representation at a reduced dimensionality of $C_{Cor}$. This schema facilitates the succeeding Transformer to invoke the self-attention mechanism in the spatial subspace while concurrently tapping into the correlation information from all SAIs ($L=WH$ and $D=C_{Cor}$).\end{highlight} The self-attention mechanism is formally defined as
\begin{equation}
    \begin{aligned}
        \label{eq:self-attention}
        \hat{I}_{Cor} & = \text{Self-Attention}(Q, K, V) \\
                      & = \text{Softmax}\left(\frac{QK^T}{\sqrt{D}}\right)V, \\
 Q,K,V & = L_Q({I_{Cor}}),L_K({I_{Cor}}),L_V({I_{Cor}})
    \end{aligned}
\end{equation}

In this context, $\hat{I}_{Cor}$ signifies the tensor enhanced by self-attention. $L_Q$, $L_K$, and $L_V$ are the linear layers for calculating queries, keys and values ($Q$, $K$ and $V$), respectively, and $D$ is their channel number. Notably, we replace commonly used predefined positional encodings with two $3 \times 3$ spatial convolutions to capture locality as suggested by \cite{chuCPVT_arxiv2021}.

\begin{highlight}
Finally, $\hat{I}_{Cor}$ undergoes the correlation decoding process to restore the angular subspace. This process mirrors the correlation encoding process in Equation \ref{eq:encoding1} and \ref{eq:encoding2}. However, it operates in reverse, using a linear layer $L_{decode}: \mathbb{R}^{B \times WH \times C_{Cor}} \mapsto \mathbb{R}^{B \times WH \times UVC}$ with a weight matrix $W_{decode} \in \mathbb{R}^{C_{Cor} \times UVC}$. The output tensor $\hat{I}$ is then generated as follows:
\begin{gather}
    \hat{I} = L_{decode}(\hat{I}_{Cor}), \\
 L_{decode}(X) = W_{decode} X.
\end{gather}
\end{highlight}

In essence, M2MT fulfills the objectives of Equation \ref{eq:after}, where $I_{Cor}$ aggregates all SAI information at each spatial location:
\begin{equation}
    I_{Cor}(x, y) \simeq \{I(\bar{u}, \bar{v}, x, y)\}_{(\bar{u}, \bar{v}) \in \mathbb{R}^{U \times V}},
\end{equation}
and the self-attention mechanism models the long-range dependencies among the spatial locations:
\begin{equation}
    \hat{I}_{Cor}(x, y) \simeq \{I_{Cor}(\bar{x}, \bar{y})\}_{(\bar{x}, \bar{y}) \in \mathbb{R}^{W \times H}} \\
\end{equation}

As a result, M2MT is enabled to access the entirety of LF data in a non-local context spatially and angularly with no information remaining isolated within the batch dimension:
\begin{equation}
    \hat{I}(u, v, x, y) \simeq \{I(\bar{u}, \bar{v}, \bar{x}, \bar{y})\}_{(\bar{u}, \bar{v}, \bar{x}, \bar{y}) \in \mathbb{R}^{U \times V \times W \times H}}
\end{equation}
where the inference process for any given location $(u, v, x, y)$ is many-to-one. Since M2MT concurrently infers all pixels, the overall operation is inherently many-to-many.

\subsection{Angular Transformer}
\hl{While M2MT achieves interactions in the spatial subspace, it remains crucial to engage an angular component to facilitate interactions within the angular subspace. To this end, an angular transformer is utilized to refine the correlation in the angular subspace.} An illustration is depicted in Fig. \ref{fig:M2MT}(c). This Transformer is fundamentally vanilla as in \cite{liangLFT_SPL2022,wangDPT_AAAI2022}, aligning closely with Equation \ref{eq:self-attention}, but specifically operates on angular tensors $I_{A} \in \mathbb{R}^{BUV \times WH \times C}$ as depicted in Fig. \ref{fig:Cubes}(c). The channel number of key, query, and value is set to $D=C$.

\hl{Notably, although the M2MT and angular Transformer operate in distinct subspaces, their primary objective converges on the establishment of a comprehensive spatial-angular representation of LF images. In Section \ref{section:ablation_altering} and TABLE \ref{tab:ablation_altering}, we demonstrate that M2MT alone establishes a solid foundation of a competitive network, incorporating angular Transformers offers a complementary effect that further enhances M2MT-Net's overall performance by effectively managing angular interactions.}
\section{Experiments}

\begin{table*}[t!]
    \caption{Quantitative comparisons with the state-of-art methods at the $2 \times$ and $4 \times$ scales across various datasets. PSNR / SSIM are used as evaluation metrics. The best and second-best results are in bold and underlined, respectively.}
    \label{tab:overall}
    \centering
    \resizebox{0.80\textwidth}{!}{
    \begin{tabular}{|l|c|c|c|c|c|c|c|}
    \hline
    Method & Scale & \textit{EPFL} &  \textit{HCInew} &  \textit{HCIold} &  \textit{INRIA} &  \textit{STFgantry} \\\hline\hline
    LFSSR \cite{yeungSAS_LFSR2019}	            & $2\times$ & 33.67/0.9744                          & 36.80/0.9749                          & 43.81/0.9938                              & 35.28/0.9832                              & 37.94/0.9898                                  \\
    LF-ATO \cite{jinLFSSRATO_2020}	            & $2\times$ & 34.27/0.9757                          & 37.24/0.9767                          & 44.21/0.9942                              & 36.17/0.9842                              & 39.64/0.9929                                  \\
    LF-InterNet \cite{wangLfInterNet_ECCV2020}	& $2\times$ & 34.11/0.9760                          & 37.17/0.9763                          & 44.57/0.9946                              & 35.83/0.9843                              & 38.44/0.9909                                  \\
    LF-IINet \cite{liuLFIINet_TMM2021}	        & $2\times$ & 34.73/0.9773                          & 37.77/0.9790                          & 44.85/0.9948                              & 36.57/0.9853                              & 39.89/0.9936                                  \\
    DKNet \cite{huDKNet_TIM2022}                & $2\times$ & 34.01/0.9759                          & 37.36/0.9780                          & 44.19/0.9942                              & 35.80/0.9843                              & 39.59/0.9910                                  \\
    DPT	\cite{wangDPT_AAAI2022}                 & $2\times$ & 34.49/0.9758                          & 37.36/0.9771                          & 44.30/0.9943                              & 36.41/0.9843                              & 39.43/0.9926                                  \\
    \hl{LFSSR-SAV \cite{chengLFSSRSAV_TCI2022}} & \hl{$2\times$} & \hl{34.62/0.9772}                & \hl{37.42/0.9776}                     & \hl{44.22/0.9942}                         & \hl{36.36/0.9849}                         & \hl{38.69/0.9914}                             \\
    DistgSSR \cite{wangDistgSSR_TIP2022}	    & $2\times$ & 34.81/0.9787                          & 37.96/0.9796                          & 44.94/0.9949                              & 36.58/0.9859                              & 40.40/0.9942                                  \\
    LFT \cite{liangLFT_SPL2022}	                & $2\times$ & 34.78/0.9776                          & 37.77/0.9788                          & 44.63/0.9947                              & 36.54/0.9853                              & 40.41/0.9941                                  \\
    EPIT \cite{liangEPIT_ICCV2023}	            & $2\times$ & 34.85/0.9775                          & 38.23/\underline{0.9810}              & 45.08/0.9949                              & 36.68/0.9852                              & \textbf{42.17}/\textbf{0.9957}                \\
    \hl{HLFSR \cite{duongHLFSR_TCI2023}}        & \hl{$2\times$} & \hl{35.31/0.9800}                & \hl{38.32/0.9807}                     & \hl{44.98/0.9950}                         & \hl{37.06/0.9867}                         & \hl{40.85/0.9947}                             \\
    \hl{LF-DET \cite{congLFDET_TMM2024}}        & \hl{$2\times$} & \hl{35.20/0.9794}                & \hl{38.22/0.9803}                     & \hl{44.92/0.9949}                         & \hl{36.88/0.9862}                         & \hl{\underline{41.56}/\underline{0.9953}}     \\
    \hline
    M2MT-Net (Ours)                             & $2\times$ & \underline{35.64}/\underline{0.9815}  & \underline{38.43}/\underline{0.9810}  & \underline{45.38}/\underline{0.9953}      & \underline{37.22}/\underline{0.9870}      & 40.99/0.9949                                  \\
    M2MT-Net* (Ours)                            & $2\times$ & \textbf{35.82}/\textbf{0.9822}        & \textbf{38.62}/\textbf{0.9816}        & \textbf{45.58}/\textbf{0.9955}            & \textbf{37.40}/\textbf{0.9873}            & 41.39/\underline{0.9953}                      \\
    \hline\hline    
    LFSSR \cite{yeungSAS_LFSR2019}              & $4\times$ & 28.60/0.9118                          & 30.93/0.9145                          & 36.91/0.9696                              & 30.59/0.9467                              & 30.57/0.9426                                  \\
    LFSSR-ATO \cite{jinLFSSRATO_2020}           & $4\times$ & 28.51/0.9115                          & 30.88/0.9135                          & 37.00/0.9699                              & 30.71/0.9484                              & 30.61/0.9430                                  \\
    LF-InterNet \cite{wangLfInterNet_ECCV2020}  & $4\times$ & 28.81/0.9162                          & 30.96/0.9161                          & 37.15/0.9716                              & 30.78/0.9491                              & 30.36/0.9409                                  \\
    LF-IINet \cite{liuLFIINet_TMM2021}          & $4\times$ & 29.04/0.9188                          & 31.33/0.9208                          & 37.62/0.9734                              & 31.03/0.9515                              & 31.26/0.9502                                  \\
    DKNet \cite{huDKNet_TIM2022}                & $4\times$ & 28.85/0.9174                          & 31.17/0.9185                          & 37.31/0.9720                              & 30.80/0.9501                              & 30.85/0.9460                                  \\
    DPT \cite{wangDPT_AAAI2022}                 & $4\times$ & 28.94/0.9170                          & 31.20/0.9188                          & 37.41/0.9721                              & 30.96/0.9503                              & 31.15/0.9488                                  \\
    \hl{LFSSR-SAV \cite{chengLFSSRSAV_TCI2022}} & \hl{$4\times$} & \hl{29.37/0.9223}                & \hl{31.45/0.9217}                     & \hl{37.50/0.9721}                         & \hl{31.27/0.9531}                         & \hl{31.36/0.9505}                             \\
    DisgSSR \cite{wangDistgSSR_TIP2022}         & $4\times$ & 28.99/0.9195                          & 31.38/0.9217                          & 37.56/0.9732                              & 30.99/0.9519                              & 31.65/0.9534                                  \\
    LFT \cite{liangLFT_SPL2022}                 & $4\times$ & 29.33/0.9196                          & 31.36/0.9205                          & 37.59/0.9731                              & 31.30/0.9515                              & 31.62/0.9548                                  \\
    EPIT \cite{liangEPIT_ICCV2023}              & $4\times$ & 29.31/0.9196                          & 31.51/0.9231                          & 37.68/0.9737                              & 31.35/0.9526                              & 32.18/0.9570                                  \\
    \hl{HLFSR \cite{duongHLFSR_TCI2023}}        & \hl{$4\times$} & \hl{29.20/0.9222}                & \hl{31.57/0.9238}                     & \hl{37.78/0.9742}                         & \hl{31.24/0.9534}                         & \hl{31.64/0.9537}                             \\
    \hl{LF-DET \cite{congLFDET_TMM2024}}        & \hl{$4\times$} & \hl{29.42/0.9220}                & \hl{31.51/0.9227}                     & \hl{37.76/0.9739}                         & \hl{31.34/0.9528}                         & \hl{32.02/0.9561}                             \\
    \hline
    M2MT-Net (Ours)                             & $4\times$ & \underline{29.85}/\underline{0.9284}  & \underline{31.76}/\underline{0.9264}  & \underline{37.98}/\underline{0.9749}      & \underline{31.77}/\underline{0.9563}      & \underline{32.20}/\underline{0.9584}          \\
    M2MT-Net* (Ours)                            & $4\times$ & \textbf{29.96}/\textbf{0.9300}        & \textbf{31.94}/\textbf{0.9279}        & \textbf{38.21}/\textbf{0.9758}            & \textbf{31.87}/\textbf{0.9572}            & \textbf{32.45}/\textbf{0.9602}                \\
    \hline
    \multicolumn{6}{l}{\scriptsize * Geometric self-ensemble strategy is applied.}
    \end{tabular}
    }
\end{table*}
\begin{figure*}[t!]
    \centering
    \tabcolsep=0.05cm
    \renewcommand{\arraystretch}{1.0}
    \resizebox{0.98\textwidth}{!}{
    \begin{tabular}{ccccccc}
        HR &
        DistgSSR &
        HLFSR &
        LFT &
        EPIT &
        LF-DET &
        M2MT-Net \\
        \hline
        \vspace{-7pt}
        \\
        \includegraphics[width=0.180\textwidth, height=0.136\textwidth]{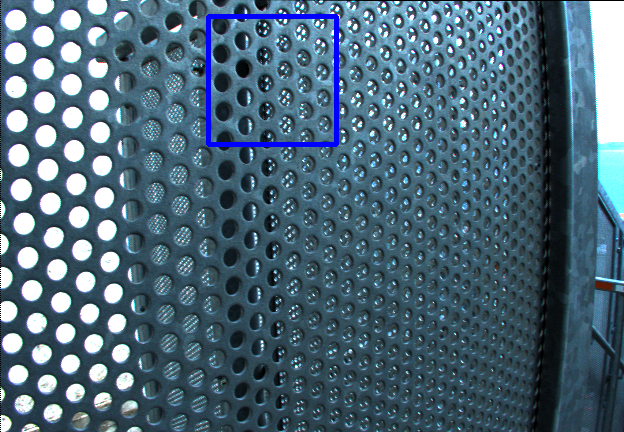} &
        \includegraphics[width=0.180\textwidth, height=0.136\textwidth]{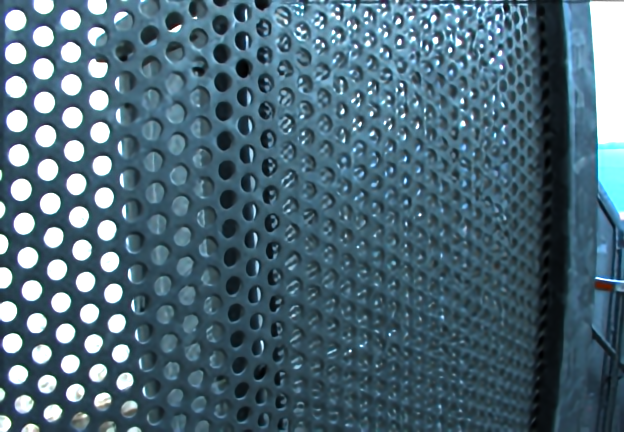} &
        \includegraphics[width=0.180\textwidth, height=0.136\textwidth]{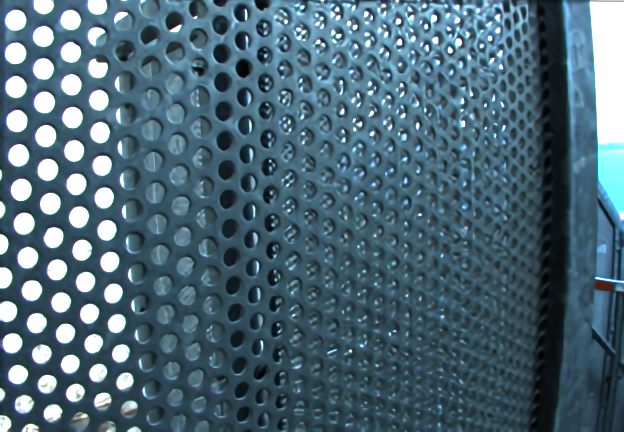} &
        \includegraphics[width=0.180\textwidth, height=0.136\textwidth]{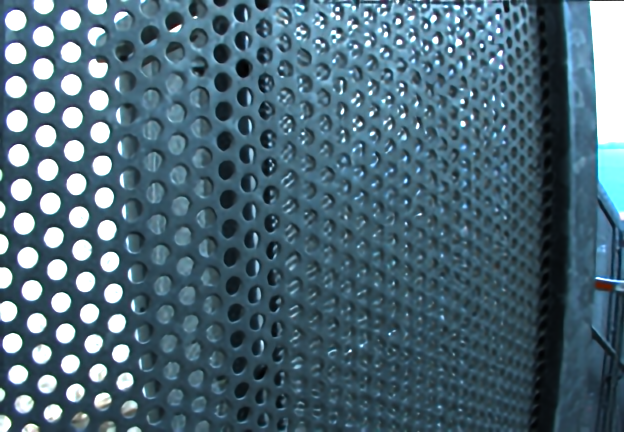} &
        \includegraphics[width=0.180\textwidth, height=0.136\textwidth]{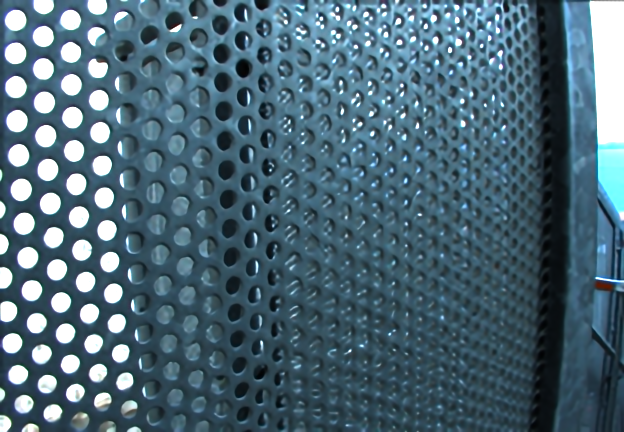} &
        \includegraphics[width=0.180\textwidth, height=0.136\textwidth]{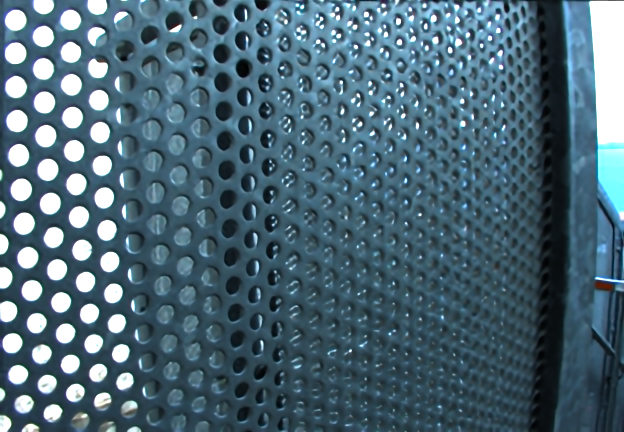} &
        \includegraphics[width=0.180\textwidth, height=0.136\textwidth]{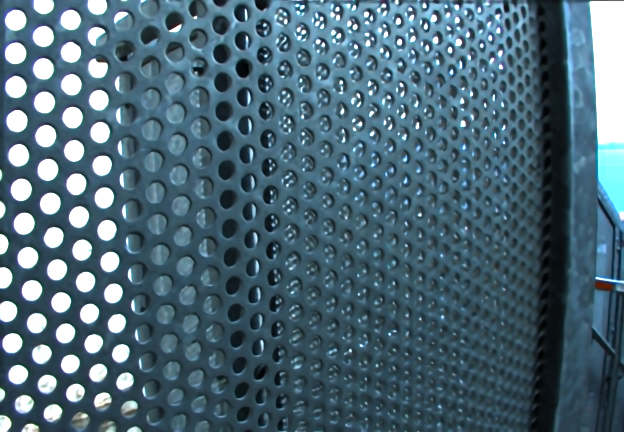} \\
        \begin{minipage}{0.180\textwidth}
            \centering
            \includegraphics[width=0.46\textwidth, height=0.46\textwidth,cfbox=blue 1pt 0pt]{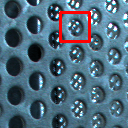}
            \includegraphics[width=0.46\textwidth, height=0.46\textwidth,cfbox=red 1pt 0pt]{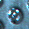}
        \end{minipage} &
        \begin{minipage}{0.180\textwidth}
            \centering
            \includegraphics[width=0.46\textwidth, height=0.46\textwidth,cfbox=blue 1pt 0pt]{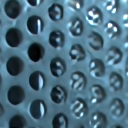}
            \includegraphics[width=0.46\textwidth, height=0.46\textwidth,cfbox=red 1pt 0pt]{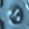}
        \end{minipage} &
        \begin{minipage}{0.180\textwidth}
            \centering
            \includegraphics[width=0.46\textwidth, height=0.46\textwidth,cfbox=blue 1pt 0pt]{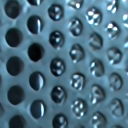}
            \includegraphics[width=0.46\textwidth, height=0.46\textwidth,cfbox=red 1pt 0pt]{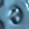}
        \end{minipage} &
        \begin{minipage}{0.180\textwidth}
            \centering
            \includegraphics[width=0.46\textwidth, height=0.46\textwidth,cfbox=blue 1pt 0pt]{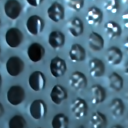}
            \includegraphics[width=0.46\textwidth, height=0.46\textwidth,cfbox=red 1pt 0pt]{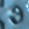}
        \end{minipage} &
        \begin{minipage}{0.180\textwidth}
            \centering
            \includegraphics[width=0.46\textwidth, height=0.46\textwidth,cfbox=blue 1pt 0pt]{img/qual/Perforated_Metal_3/EPIT/SR.1.png}
            \includegraphics[width=0.46\textwidth, height=0.46\textwidth,cfbox=red 1pt 0pt]{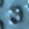}
        \end{minipage} &
        \begin{minipage}{0.180\textwidth}
            \centering
            \includegraphics[width=0.46\textwidth, height=0.46\textwidth,cfbox=blue 1pt 0pt]{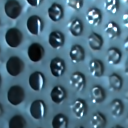}
            \includegraphics[width=0.46\textwidth, height=0.46\textwidth,cfbox=red 1pt 0pt]{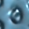}
        \end{minipage} &
        \begin{minipage}{0.180\textwidth}
            \centering
            \includegraphics[width=0.46\textwidth, height=0.46\textwidth,cfbox=blue 1pt 0pt]{img/qual/Perforated_Metal_3/M2MTNet/SR.1.png}
            \includegraphics[width=0.46\textwidth, height=0.46\textwidth,cfbox=red 1pt 0pt]{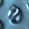}
        \end{minipage} \\
        (a) \textit{Perforated\_Metal\_3} &
        24.14/0.7053 &
        24.57/0.7216 &
        24.24/0.7040 &
        24.18/0.6989 &
        \underline{24.81}/\underline{0.7218} &
        \textbf{25.49}/\textbf{0.7573} \\
        \vspace{-10pt}
        \\
        \raisebox{1.8\height}{
        \resizebox{0.06\textwidth}{!}{
            \begin{tikzpicture}
                \foreach \x in {0,1,2,3,4} {
                    \foreach \y in {0,1,2,3,4} {
                        \draw[black, thin] (\x,\y) rectangle (\x+1,\y+1);
                    }
                }
                \fill[red] (2,2) rectangle (3,3);
            \end{tikzpicture}
        } } &
        \imageWithGrid{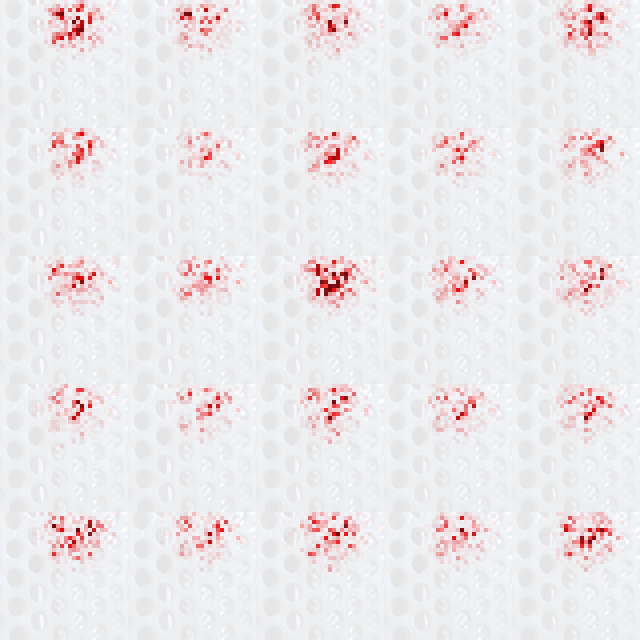}{0.178\textwidth}{0.178\textwidth} &
        \imageWithGrid{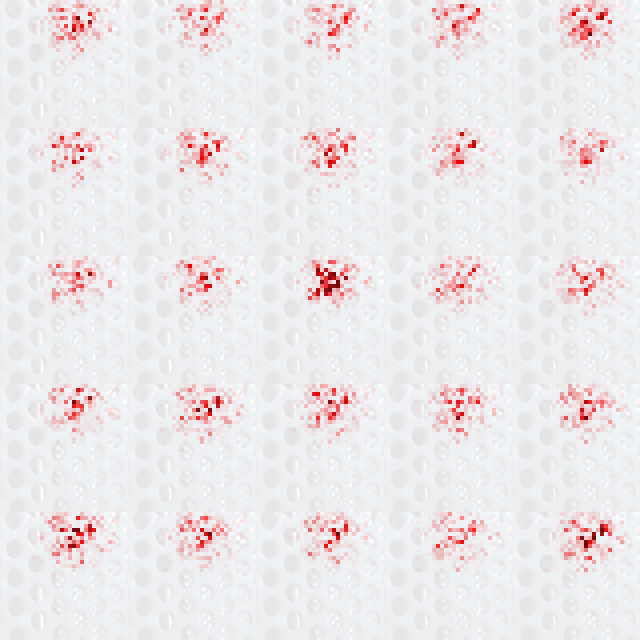}{0.178\textwidth}{0.178\textwidth} &
        \imageWithGrid{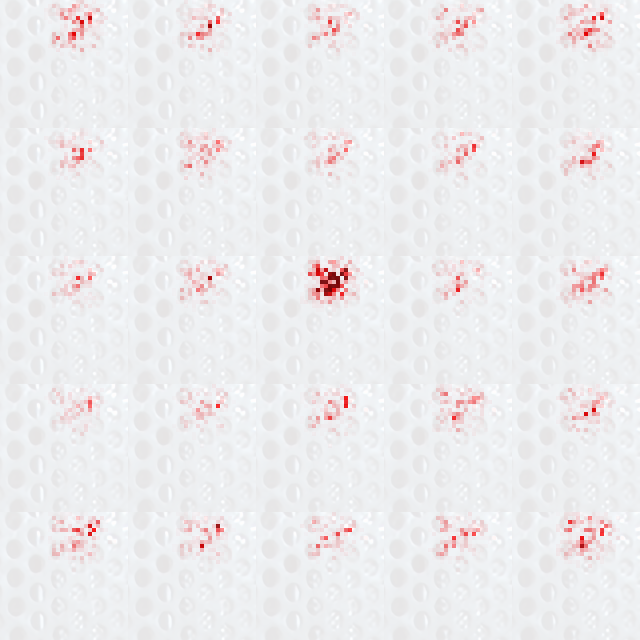}{0.178\textwidth}{0.178\textwidth} &
        \imageWithGrid{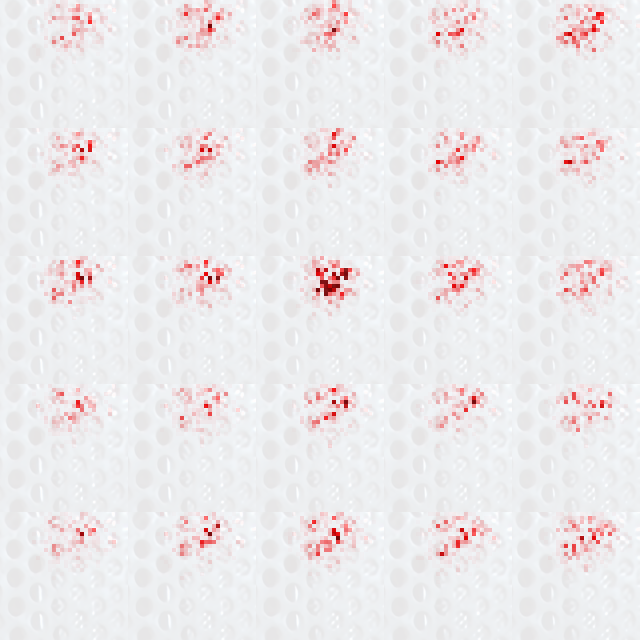}{0.178\textwidth}{0.178\textwidth} &
        \imageWithGrid{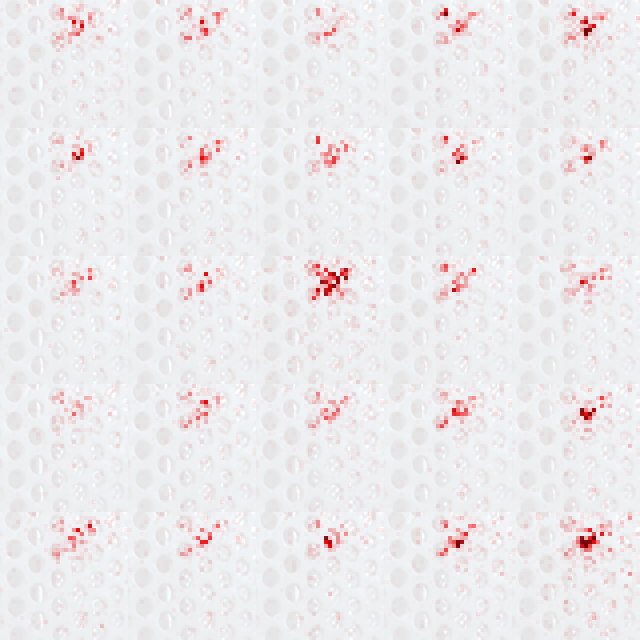}{0.178\textwidth}{0.178\textwidth} &
        \imageWithGrid{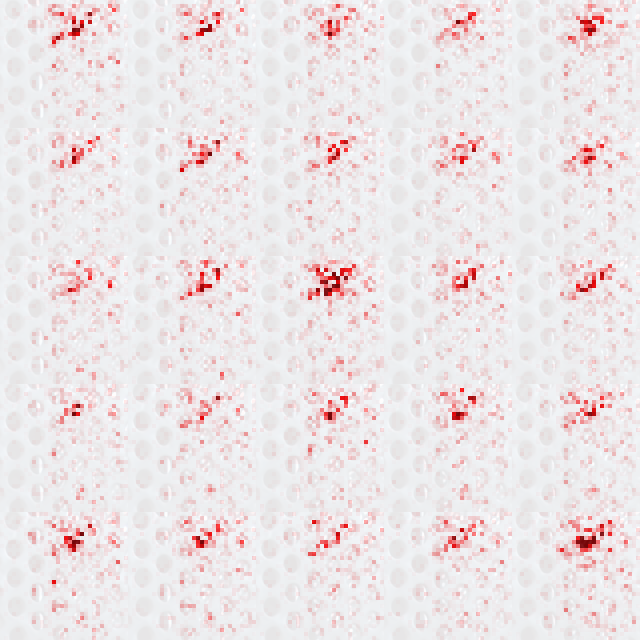}{0.178\textwidth}{0.178\textwidth} \\
        &
        DI = 8.7671 &
        DI = 12.5190 &
        DI = 5.8032 &
        DI = 8.3895 &
        DI = 23.3273 &
        DI = 25.1975 \\\hline

        \vspace{-7pt}
        \\

        \includegraphics[width=0.180\textwidth, height=0.136\textwidth]{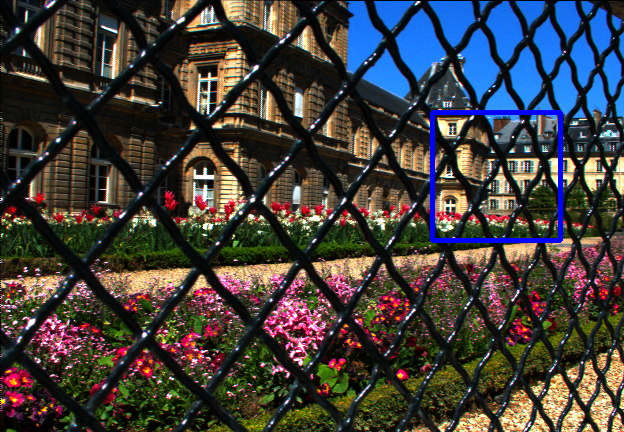} &
        \includegraphics[width=0.180\textwidth, height=0.136\textwidth]{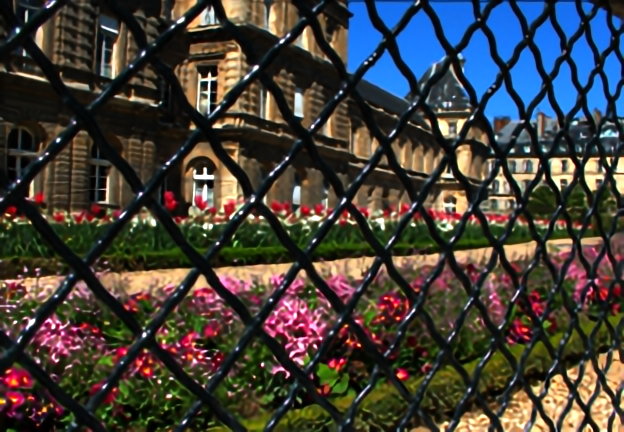} &
        \includegraphics[width=0.180\textwidth, height=0.136\textwidth]{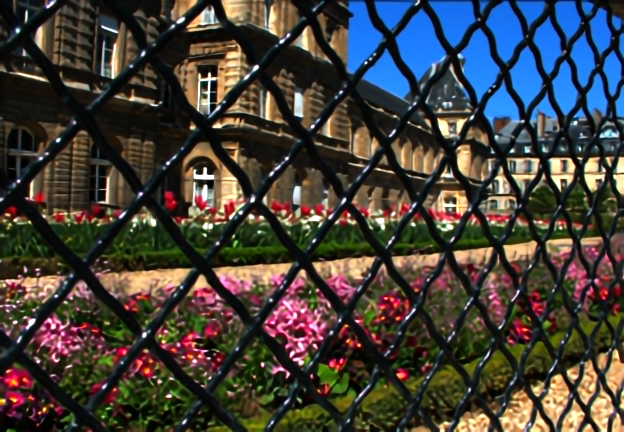} &
        \includegraphics[width=0.180\textwidth, height=0.136\textwidth]{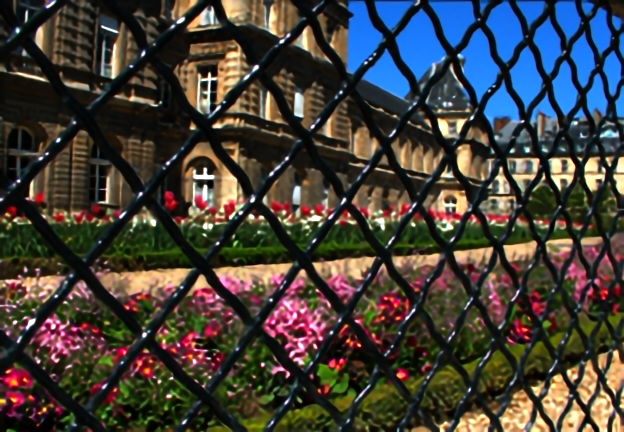} &
        \includegraphics[width=0.180\textwidth, height=0.136\textwidth]{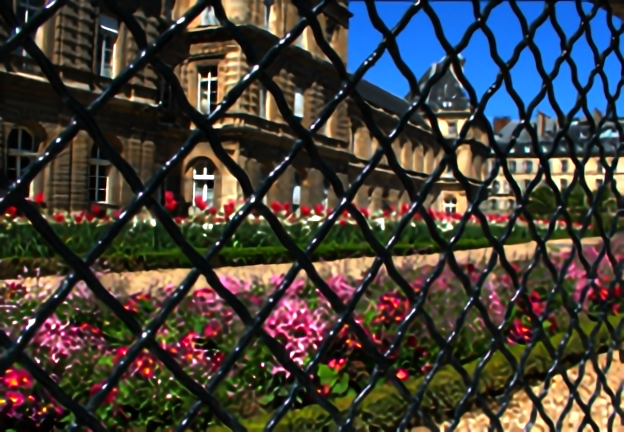} &
        \includegraphics[width=0.180\textwidth, height=0.136\textwidth]{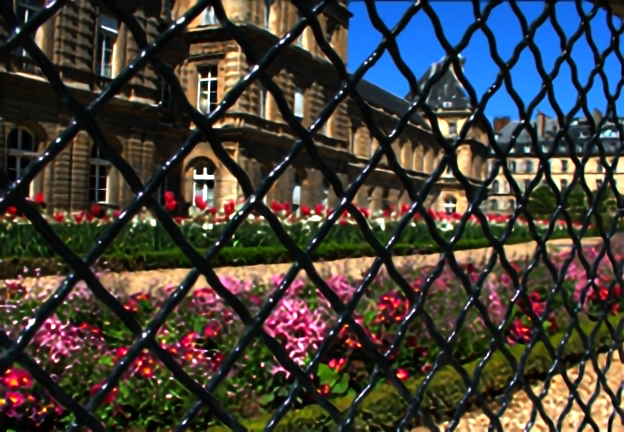} &
        \includegraphics[width=0.180\textwidth, height=0.136\textwidth]{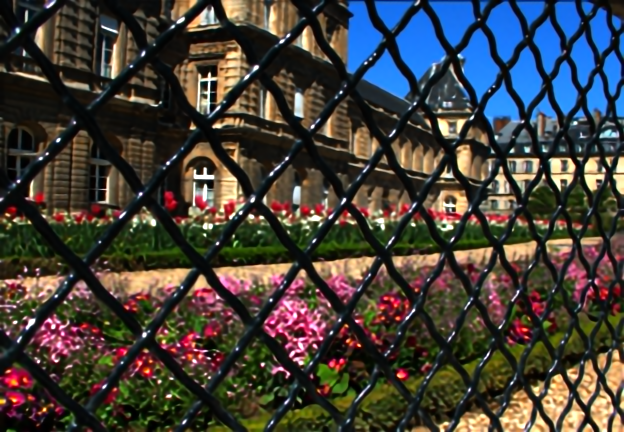} \\
        \begin{minipage}{0.180\textwidth}
            \centering
            \includegraphics[width=0.46\textwidth, height=0.46\textwidth,cfbox=blue 1pt 0pt]{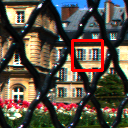}
            \includegraphics[width=0.46\textwidth, height=0.46\textwidth,cfbox=red 1pt 0pt]{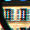}
        \end{minipage} &
        \begin{minipage}{0.180\textwidth}
            \centering
            \includegraphics[width=0.46\textwidth, height=0.46\textwidth,cfbox=blue 1pt 0pt]{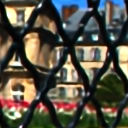}
            \includegraphics[width=0.46\textwidth, height=0.46\textwidth,cfbox=red 1pt 0pt]{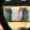}
        \end{minipage} &
        \begin{minipage}{0.180\textwidth}
            \centering
            \includegraphics[width=0.46\textwidth, height=0.46\textwidth,cfbox=blue 1pt 0pt]{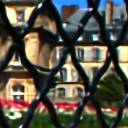}
            \includegraphics[width=0.46\textwidth, height=0.46\textwidth,cfbox=red 1pt 0pt]{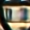}
        \end{minipage} &
        \begin{minipage}{0.180\textwidth}
            \centering
            \includegraphics[width=0.46\textwidth, height=0.46\textwidth,cfbox=blue 1pt 0pt]{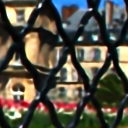}
            \includegraphics[width=0.46\textwidth, height=0.46\textwidth,cfbox=red 1pt 0pt]{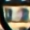}
        \end{minipage} &
        \begin{minipage}{0.180\textwidth}
            \centering
            \includegraphics[width=0.46\textwidth, height=0.46\textwidth,cfbox=blue 1pt 0pt]{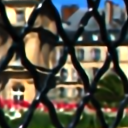}
            \includegraphics[width=0.46\textwidth, height=0.46\textwidth,cfbox=red 1pt 0pt]{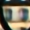}
        \end{minipage} &
        \begin{minipage}{0.180\textwidth}
            \centering
            \includegraphics[width=0.46\textwidth, height=0.46\textwidth,cfbox=blue 1pt 0pt]{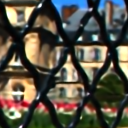}
            \includegraphics[width=0.46\textwidth, height=0.46\textwidth,cfbox=red 1pt 0pt]{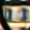}
        \end{minipage} &
        \begin{minipage}{0.180\textwidth}
            \centering
            \includegraphics[width=0.46\textwidth, height=0.46\textwidth,cfbox=blue 1pt 0pt]{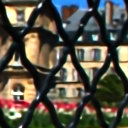}
            \includegraphics[width=0.46\textwidth, height=0.46\textwidth,cfbox=red 1pt 0pt]{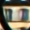}
        \end{minipage} \\
        (b) \textit{Palais\_du\_Luxembourg} &
        22.69/0.7237 &
        22.75/0.7295 &
        22.70/0.7218 &
        22.58/0.7115 &
        \underline{22.87}/\underline{0.7325} &
        \textbf{23.16}/\textbf{0.7466} \\
        \vspace{-10pt}
        \\
        \raisebox{1.8\height}{
        \resizebox{0.06\textwidth}{!}{
            \begin{tikzpicture}
                \foreach \x in {0,1,2,3,4} {
                    \foreach \y in {0,1,2,3,4} {
                        \draw[black, thin] (\x,\y) rectangle (\x+1,\y+1);
                    }
                }
                \fill[red] (1,1) rectangle (2,2);
            \end{tikzpicture}
        } } &
        \imageWithGrid{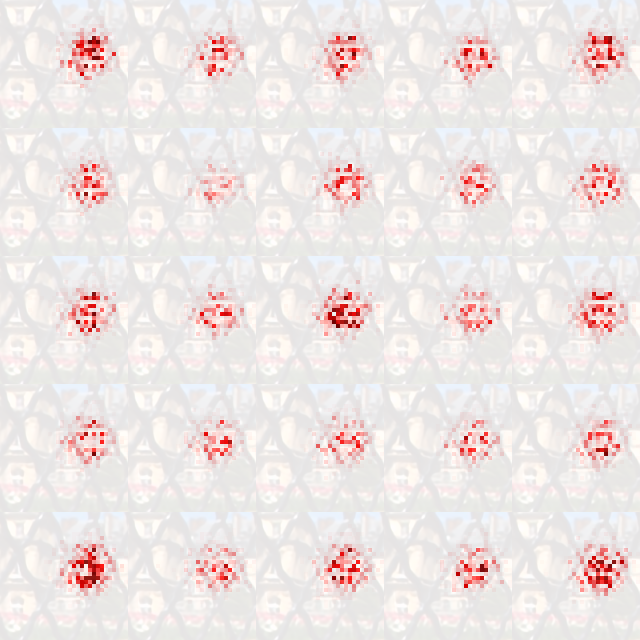}{0.178\textwidth}{0.178\textwidth} &
        \imageWithGrid{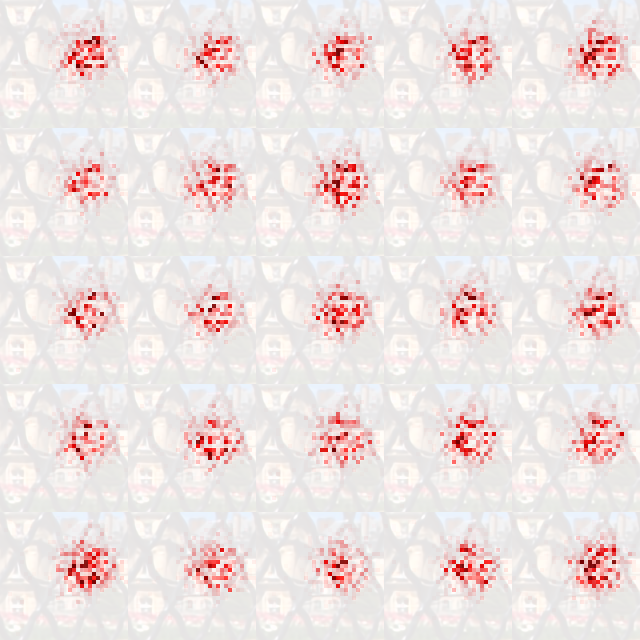}{0.178\textwidth}{0.178\textwidth} &
        \imageWithGrid{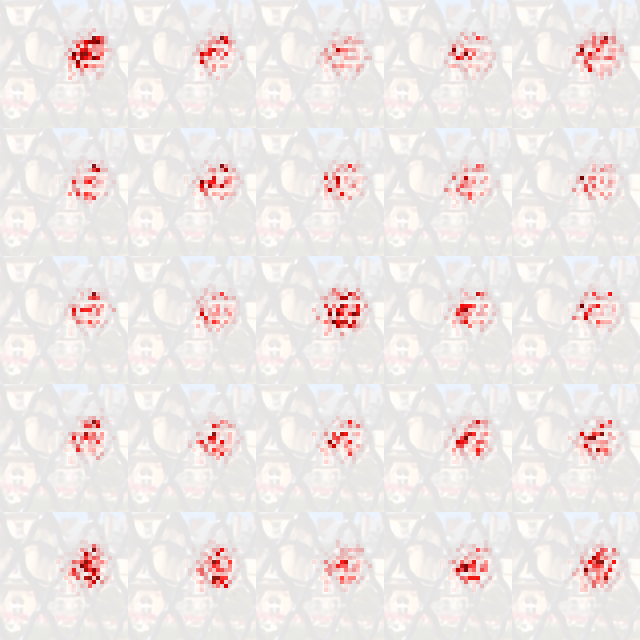}{0.178\textwidth}{0.178\textwidth} &
        \imageWithGrid{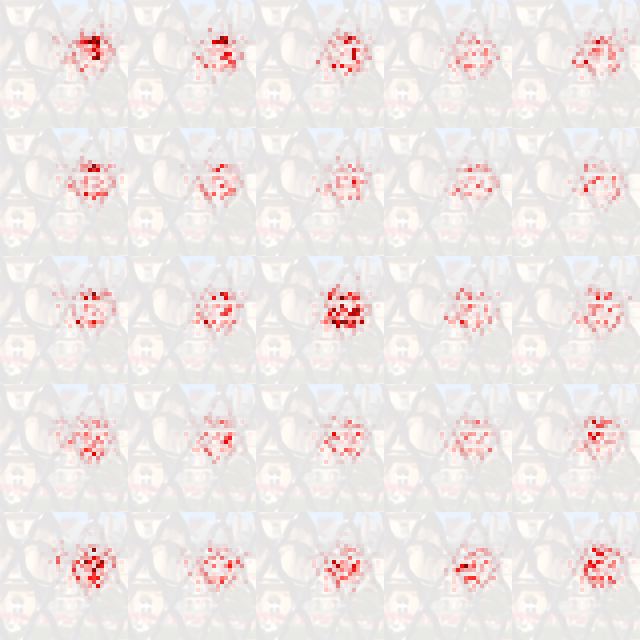}{0.178\textwidth}{0.178\textwidth} &
        \imageWithGrid{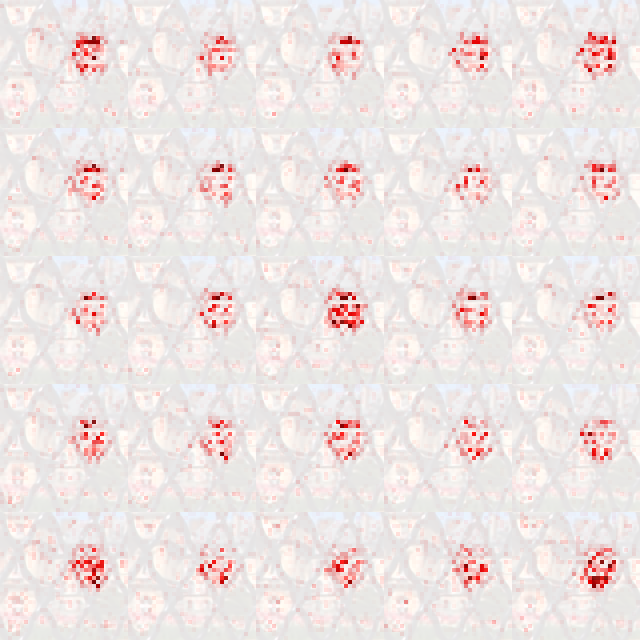}{0.178\textwidth}{0.178\textwidth} &
        \imageWithGrid{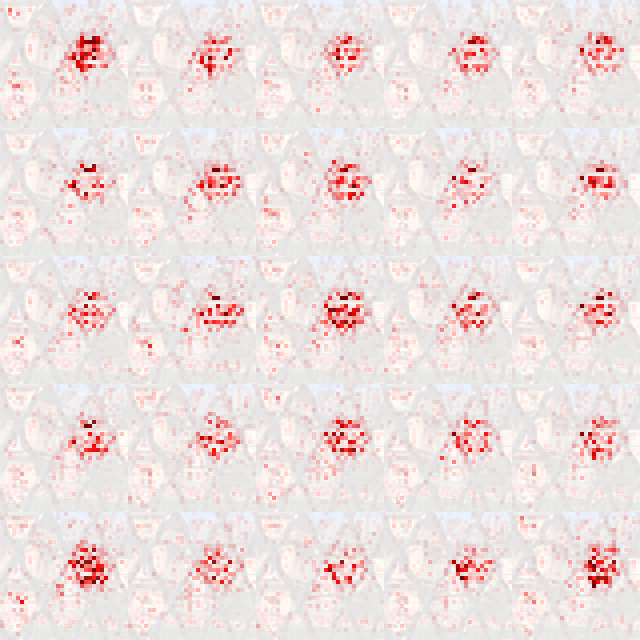}{0.178\textwidth}{0.178\textwidth} \\
        &
        DI = 8.0693 &
        DI = 12.5424 &
        DI = 5.1459 &
        DI = 8.2841 &
        DI = 23.7106 &
        DI = 25.2688 \\\hline

        \vspace{-7pt}
        \\

        \includegraphics[width=0.180\textwidth, height=0.160\textwidth]{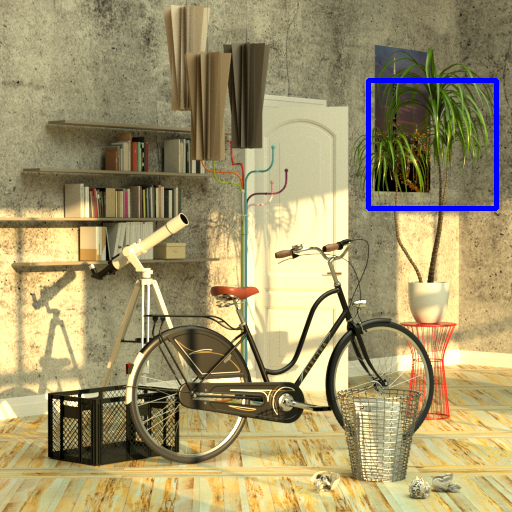} &
        \includegraphics[width=0.180\textwidth, height=0.160\textwidth]{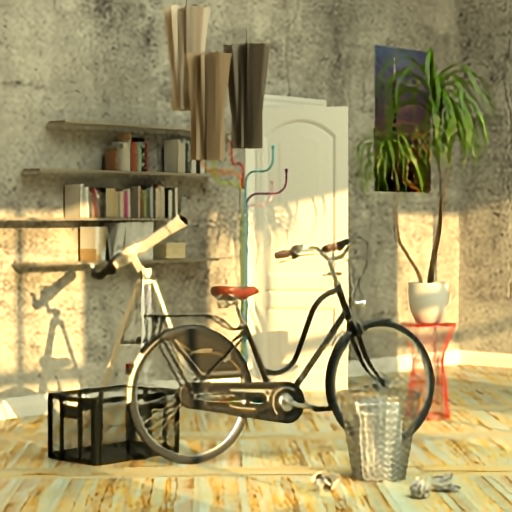} &
        \includegraphics[width=0.180\textwidth, height=0.160\textwidth]{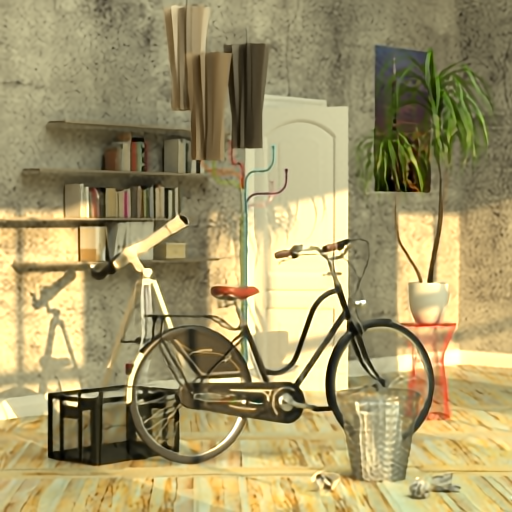} &
        \includegraphics[width=0.180\textwidth, height=0.160\textwidth]{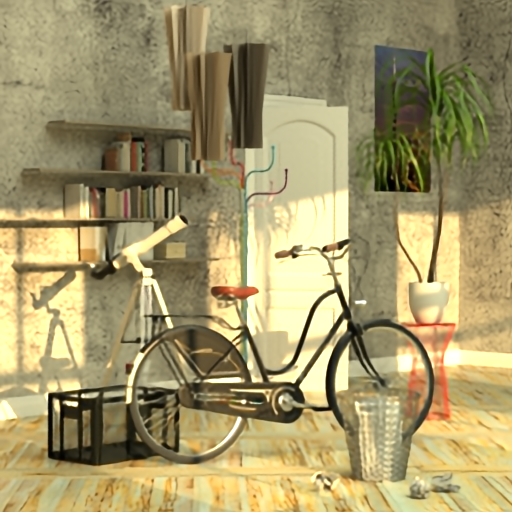} &
        \includegraphics[width=0.180\textwidth, height=0.160\textwidth]{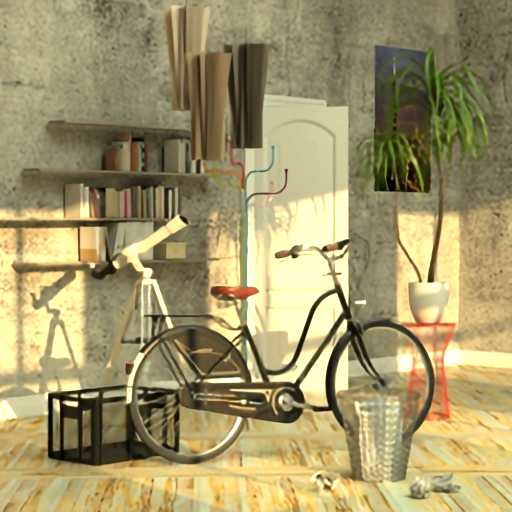} &
        \includegraphics[width=0.180\textwidth, height=0.160\textwidth]{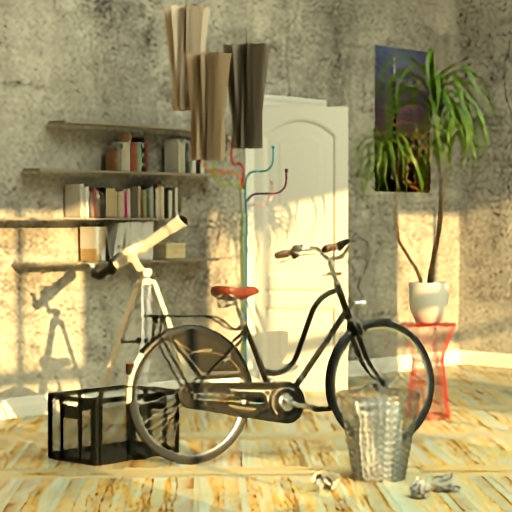} &
        \includegraphics[width=0.180\textwidth, height=0.160\textwidth]{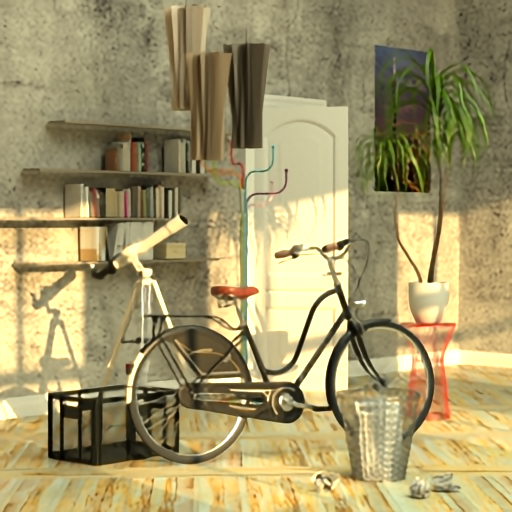} \\
        \begin{minipage}{0.180\textwidth}
            \centering
            \includegraphics[width=0.46\textwidth, height=0.46\textwidth,cfbox=blue 1pt 0pt]{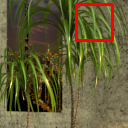}
            \includegraphics[width=0.46\textwidth, height=0.46\textwidth,cfbox=red 1pt 0pt]{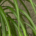}
        \end{minipage} &
        \begin{minipage}{0.180\textwidth}
            \centering
            \includegraphics[width=0.46\textwidth, height=0.46\textwidth,cfbox=blue 1pt 0pt]{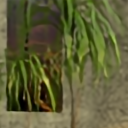}
            \includegraphics[width=0.46\textwidth, height=0.46\textwidth,cfbox=red 1pt 0pt]{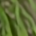}
        \end{minipage} &
        \begin{minipage}{0.180\textwidth}
            \centering
            \includegraphics[width=0.46\textwidth, height=0.46\textwidth,cfbox=blue 1pt 0pt]{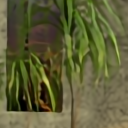}
            \includegraphics[width=0.46\textwidth, height=0.46\textwidth,cfbox=red 1pt 0pt]{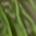}
        \end{minipage} &
        \begin{minipage}{0.180\textwidth}
            \centering
            \includegraphics[width=0.46\textwidth, height=0.46\textwidth,cfbox=blue 1pt 0pt]{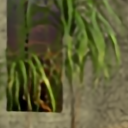}
            \includegraphics[width=0.46\textwidth, height=0.46\textwidth,cfbox=red 1pt 0pt]{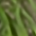}
        \end{minipage} &
        \begin{minipage}{0.180\textwidth}
            \centering
            \includegraphics[width=0.46\textwidth, height=0.46\textwidth,cfbox=blue 1pt 0pt]{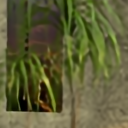}
            \includegraphics[width=0.46\textwidth, height=0.46\textwidth,cfbox=red 1pt 0pt]{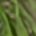}
        \end{minipage} &
        \begin{minipage}{0.180\textwidth}
            \centering
            \includegraphics[width=0.46\textwidth, height=0.46\textwidth,cfbox=blue 1pt 0pt]{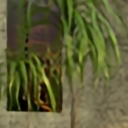}
            \includegraphics[width=0.46\textwidth, height=0.46\textwidth,cfbox=red 1pt 0pt]{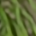}
        \end{minipage} &
        \begin{minipage}{0.180\textwidth}
            \centering
            \includegraphics[width=0.46\textwidth, height=0.46\textwidth,cfbox=blue 1pt 0pt]{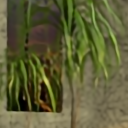}
            \includegraphics[width=0.46\textwidth, height=0.46\textwidth,cfbox=red 1pt 0pt]{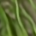}
        \end{minipage} \\
        (c) \textit{Bicycle} &
        26.32/0.6822 &
        26.54/\underline{0.6913} &
        26.33/0.6806 &
        26.46/0.6858 &
        \underline{26.55}/0.6885 &
        \textbf{26.85}/\textbf{0.7001} \\
        \vspace{-10pt}
        \\
        \raisebox{1.8\height}{
        \resizebox{0.06\textwidth}{!}{
            \begin{tikzpicture}
                \foreach \x in {0,1,2,3,4} {
                    \foreach \y in {0,1,2,3,4} {
                        \draw[black, thin] (\x,\y) rectangle (\x+1,\y+1);
                    }
                }
                \fill[red] (2,2) rectangle (3,3);
            \end{tikzpicture}
        } } &
        \imageWithGrid{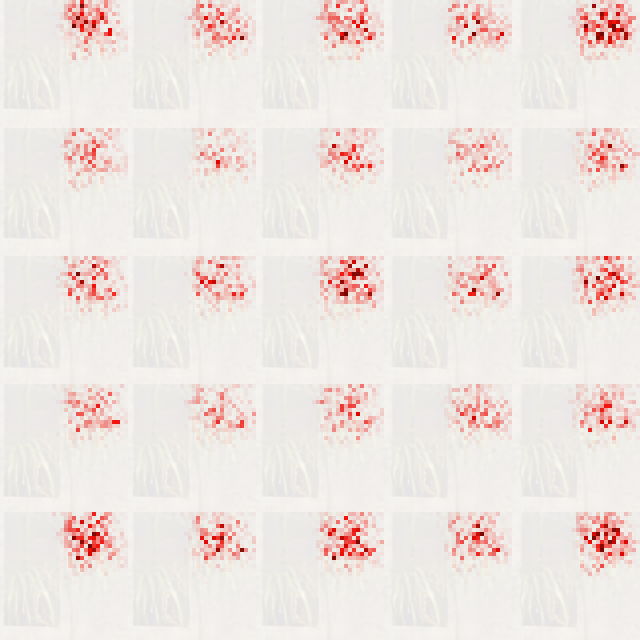}{0.178\textwidth}{0.178\textwidth} &
        \imageWithGrid{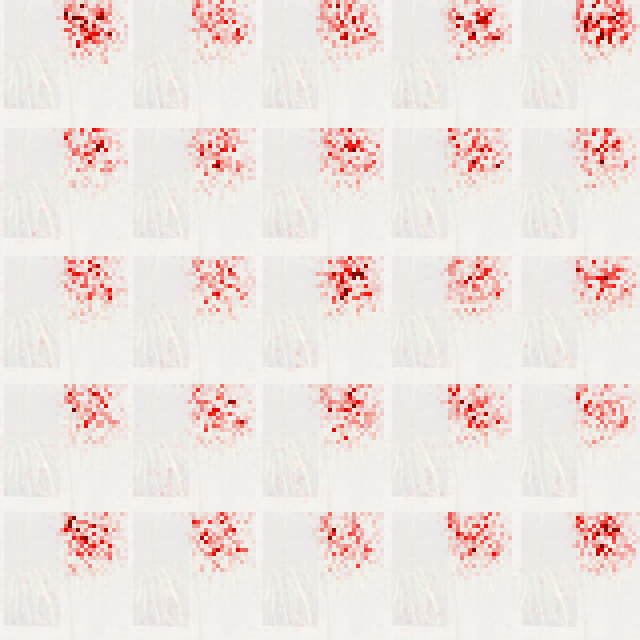}{0.178\textwidth}{0.178\textwidth} &
        \imageWithGrid{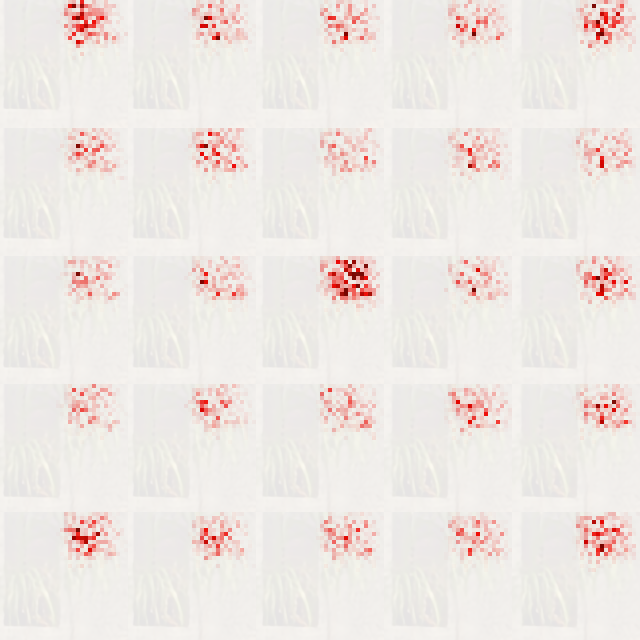}{0.178\textwidth}{0.178\textwidth} &
        \imageWithGrid{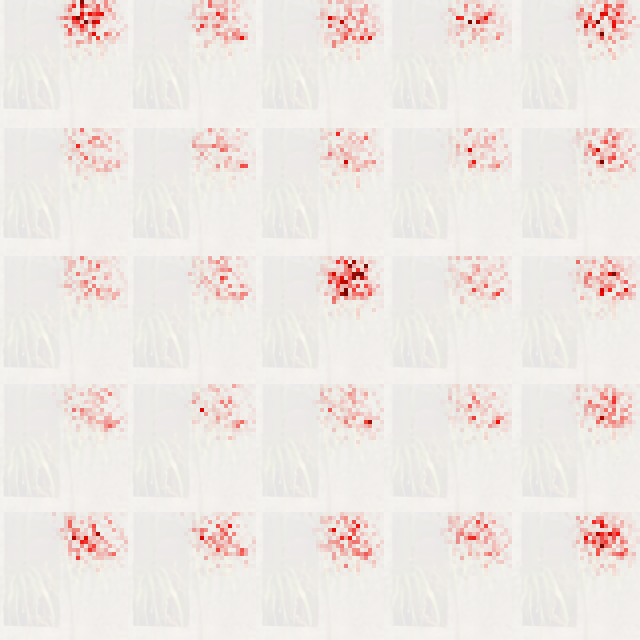}{0.178\textwidth}{0.178\textwidth} &
        \imageWithGrid{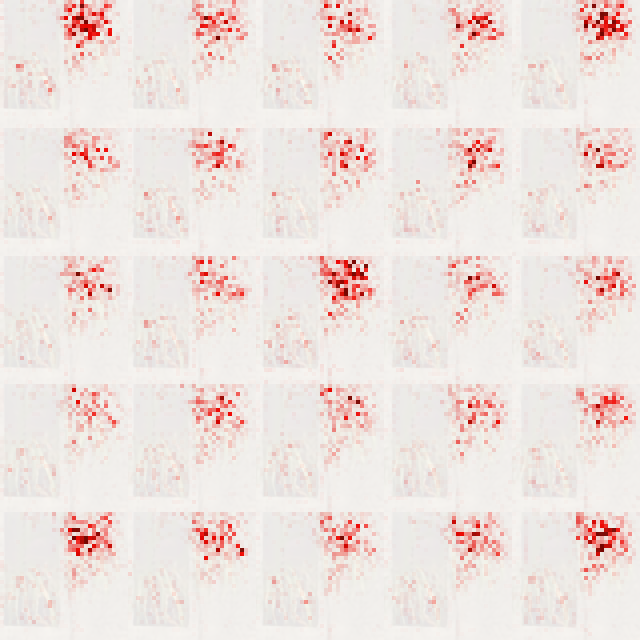}{0.178\textwidth}{0.178\textwidth} &
        \imageWithGrid{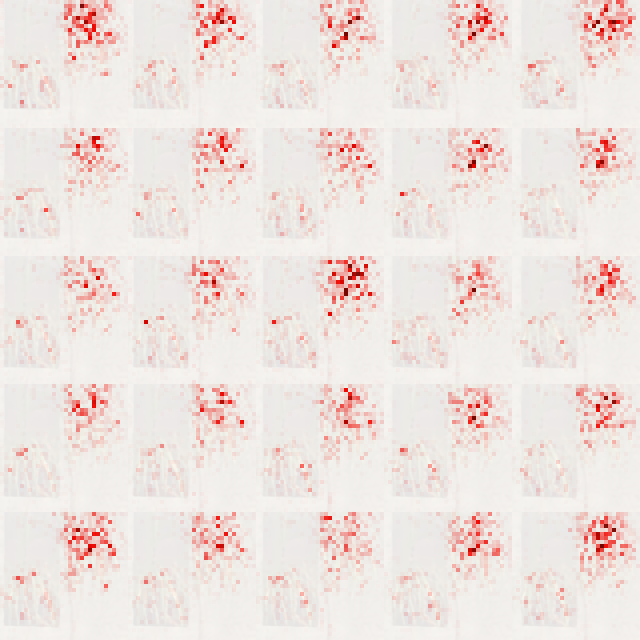}{0.178\textwidth}{0.178\textwidth} \\
        &
        DI = 9.1141 &
        DI = 13.1903 &
        DI = 6.6828 &
        DI = 8.2787 &
        DI = 18.7618 &
        DI = 19.7059 \\
    \end{tabular}
    }

    \caption{Visualization of selected samples in the $4\times$ task. In each sample, the following result is provided for each compared method: the SAI, the zoom-in views from the blue and red boxes, the PSNR/SSIM of the red box, the Local Attribution Map (LAM) of the red box and its Diffusion Index (DI). The best and second-best PSNR/SSIM are in bold and underlined. The angular location indicator is given below the HR.}
    \label{fig:Qual}
\end{figure*}
\begin{figure*}[t!]
    \centering
    \tabcolsep=0.05cm
    \renewcommand{\arraystretch}{1.0}
    \begin{tabular}{c}
        \includegraphics[width=0.98\textwidth]{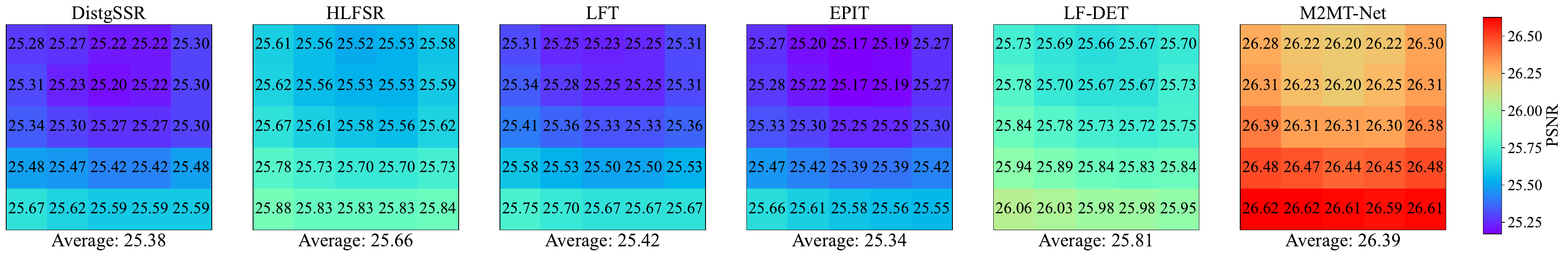} \\
        (a) \textit{Perforated\_Metal\_3} \\
        \includegraphics[width=0.98\textwidth]{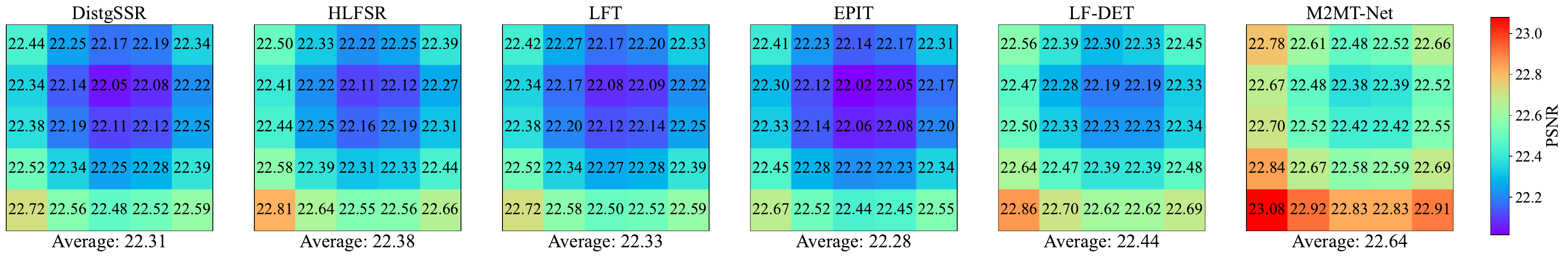} \\
        (b) \textit{Palais\_du\_Luxembourg} \\
        \includegraphics[width=0.98\textwidth]{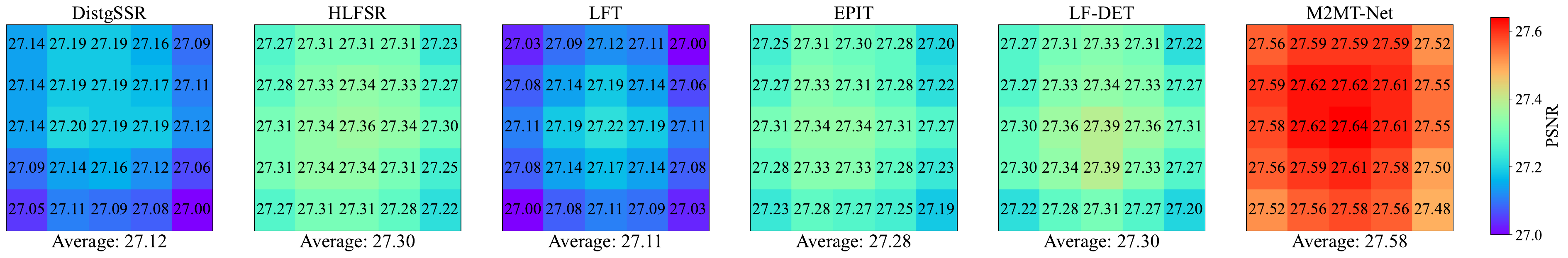} \\
        (c) \textit{Bicycle}
    \end{tabular}

    \caption{Visualization of SAI-wise PSNR to demonstrate the distribution of $4\times$ LFSR performance. The compared samples are the same with Fig. \ref*{fig:Qual}.}
    \label{fig:QualMatrix}
\end{figure*}

\subsection{Implementation Details}

In our experiments, M2MT-Net is implemented using the deep learning framework PyTorch \cite{PyTorch}. We adhere to the protocols outlined in the widely used BasicLFSR framework \cite{BasicLFSR} to conduct evaluations in a fair and consistent manner. Five public datasets are used, namely \textit{EPFL} \cite{rerabekEPFL2016}, \textit{HCInew} \cite{honauerHCInew_ACCV2016}, \textit{HCIold} \cite{wannerHCIold_VMV2013}, \textit{INRIA} \cite{lependuINRIA_TIP2018}, and \textit{STFgantry} \cite{vaishSTFgantry_2008}. These datasets contain 70/20/10/35/9 samples for training and 10/4/2/5/2 samples for testing. Following the protocols, we only use the central $5 \times 5$ SAIs \cite{BasicLFSR}. For training, each SAI was partitioned into $64 \times 64$ or $128 \times 128$ patches to serve as HR patches, and $1/2$ or $1/4$ bicubic down-sampling is applied to produce the corresponding LR patches for $2 \times$ or $4 \times$ scales, respectively. We use Adam optimizer with a learning rate of $2 \times 10^{-4}$ and batches of 4 samples. The training process takes 60 epochs to converge and five epochs to fine-tune. Regarding the hyperparameters, empirically, we use $C=48$ across all Transformers and convolutions except M2MT's correlation tensors and query, key and values with $C_{Cor} = D = 128$. \hl{The number of spatial convolutions in the initial feature extraction $n_1$ is set to 4. The number of correlation blocks $n_2$ is set to 9 for the $2 \times$ scale and 8 for the $4 \times$ scale.}

The experiments are conducted on a computer with an Intel i7-11700 4.800GHz 8-core CPU, 32 MB RAM, and an Nvidia GTX 3090 GPU. The implementation code and trained models are released publicly at \href{https://huzexi.github.io/}{https://huzexi.github.io/}.

\subsection{Quantitative Comparisons}
A quantitative comparison is conducted to compare M2MT-Net with eight state-of-the-art LFSR methods on the five aforementioned datasets at the $2 \times$ and $4 \times$ scales. The compared methods include convolution-based LFSSR \cite{yeungSAS_LFSR2019}, LF-ATO \cite{jinLFSSRATO_2020}, LF-InterNet \cite{wangLfInterNet_ECCV2020}, LF-IINet \cite{liuLFIINet_TMM2021}, DKNet \cite{huDKNet_TIM2022}, \hl{LFSSR-SAV \cite{chengLFSSRSAV_TCI2022}}, DistgSSR \cite{wangDistgSSR_TIP2022}, \hl{HLFSR \cite{duongHLFSR_TCI2023}} and Transformer-based DPT \cite{wangDPT_AAAI2022}, LFT \cite{liangLFT_SPL2022}, EPIT \cite{liangEPIT_ICCV2023} and \hl{LF-DET \cite{congLFDET_TMM2024}}. Their publicly released weight files are utilized to conduct this comparison. The outcomes are presented in Table \ref{tab:overall}.

It is evident that our M2MT-Net holds a superior position. At both the $2 \times$ and $4 \times$ scales, it achieves the highest PSNR across almost all datasets. Notably, on the \textit{EPFL} dataset, which contains the most testing samples, M2MT-Net surpasses the second-best method, LF-DET, by a significant 0.43 dB PSNR gain at the $4 \times$ scale and 0.33 dB at the $2 \times$ scale. M2MT-Net ranks third on only one dataset, \textit{STFgantry}, at the $2 \times scale$. This particular outcome can be attributed to the dataset's distinctive characteristic of exhibiting high disparities, which is effectively addressed by the EPI mechanism of EPIT and HLFSR and LF-DET's multi-scale angular modeling. However, their advantage does not extend to the $4 \times$ scale, where M2MT-Net reclaims its lead, surpassing EPIT, HLFSR and LF-DET by margins of 0.02 dB, 0.56 dB and 0.18 dB, respectively.

A notable trend is observed regarding the performance between the $2 \times$ and $4 \times$ scales. While the PSNR advantage of M2MT-Net over the second best methods at $2 \times$ scale is relatively modest from 0.11 to 0.30 dB, the gap widens significantly at the $4 \times$ scale, ranging from 0.19 to 0.43 dB. This discrepancy highlights the inherent strengths of M2MT-Net in handling the more challenging $4 \times$ scale, where more details are lost due to down-sampling, requiring the model to utilize existing spatial and angular cues more effectively.

We also incorporate the geometric self-ensemble strategy, which was initially proposed for single image super-resolution \cite{limEDSR_CVPRW2017}, into M2MT-Net to enhance the model performance without introducing additional parameters. The variant is labeled as M2MT-Net* in Table \ref{tab:overall}. Similar to its application in 2D single images, during inference, the strategy transforms the 2D LR by flipping and rotating to construct an ensemble $\{T_{i}(\overline{I}_{LR})\}$, where $T_{i}$ represents a transform function. The SR is generated by executing the network on each member in the ensemble individually, followed by the corresponding inverse transform, and finally, averaging the output. The strategy is expressed as
\begin{equation}
 I_{SR} = \frac{1}{n} \sum_{i = 1}^{n} T_{i}^{-1}(\mathcal{F}(T_{i}(I_{LR})))
\end{equation}
where $n$ is the number of transforms. The transforms take place on the spatial and angular subspaces synergistically to ensure that the LF structure is not distorted but preserved after the transforms. The result in Table \ref{tab:overall} demonstrates the advantageous impact brought by the strategy with a roughly 0.10 to 0.25 dB increase in PSNR observed across the datasets at both scales and a particular 0.40 dB increase on the \textit{STFgantry} dataset at the $2 \times$ scales respectively. These findings suggest that the geometric self-ensemble strategy is a valuable addition to compensate for LFSR models.

\subsection{Qualitative Comparisons}
We further explore the superior performance of M2MT-Net in qualitative evaluation. Fig. \ref{fig:Qual} presents qualitative results at the $4 \times$ scale for three representative samples, namely (a) \textit{Perforated\_Metal\_3}, (b) \textit{Palais\_du\_Luxembourg} and (c) \textit{Bicycle}. The first two samples are from the \textit{EPFL} dataset captured by Lytro cameras \cite{Lytro}, and the third one is from the synthetic dataset \textit{HCInew}. We compared M2MT-Net with five methods: DistgSSR and HLFSR represent the convolution-based methods, while LFT, EPIT \hl{and LF-DET} represent the Transformer-based methods. Zoom-in views inside blue and red boxes are provided to show more details. Accompanying these visuals, PSNR and SSIM are calculated on the red box areas. In general, all these techniques capably enhance resolution and preserve primary structures, but nuanced distinctions emerge within the details, especially the zoom-in views of red boxes.

In the \textit{Perforated\_Metal\_3} sample, most methods portray the perforated hole reasonably well but fall short in edge sharpness, likely influenced by lighting and occlusion challenges. M2MT-Net, however, produces notably sharper edges and a more round shape of the hole. In the \textit{Palais\_du\_Luxembourg} sample, M2MT-Net excels in reconstructing the edges of windows beyond the other methods. For the \textit{Bicycle} sample, the edges of the leaves are clearly sharper in M2MT-Net than others.

In Fig. \ref{fig:QualMatrix}, SAI-wise PSNR on these three samples is visualized. The visual representation highlights M2MT-Net's notable enhancements across SAIs. Notably, in cases (a) \textit{Perforated\_Metal\_3} and (c) \textit{Bicycle}, the lowest PSNR values achieved by M2MT-Net are still higher than the highest PSNR values of other methods. This observation signifies M2MT-Net's consistent superiority across SAIs.

\subsection{LAM Analysis}
\begin{highlight}
To further probe into the underlying capability of M2MT-Net, we employ the Local Attribution Map (LAM) technique \cite{guLAM_CVPR2021}, an attribution approach to identify pixels in the input that have a significant impact on the generation of a target window in the output, to provide insight and transparency into the performance of M2MT-Net.

Assuming $\mathcal{F}(\cdot)$ is the super-resolution network as stated before and $\mathcal{D}(\cdot)$ is a detector of edges and textures, with the detector operating on the super-resolved result as $\mathcal{D}(\mathcal{F}(\cdot))$, the LAM is derived by calculating its path integrated gradient along a gradually changing path function $\gamma(\cdot)$ as follows:
\begin{flalign}
    \begin{aligned}
    &LAM_{\mathcal{F},\mathcal{D}}(\gamma)_i := \\
    &\sum_{k=1}^{m}
    \frac{ \partial \mathcal{D}(\mathcal{F}(\gamma(\frac{k}{m}))) }{\partial \gamma(\frac{k}{m})_i} \cdot 
 (\gamma(\frac{k}{m}) - \gamma(\frac{k+1}{m}))_i \cdot
    \frac{1}{m}
    \end{aligned}
\end{flalign}
where $i$ is the dimension index, and $m$ and $k$ are the number of steps and the step index in the path, respectively. $k$ is set to 50 in the analysis. Here, the detector $\mathcal{D}$ is a simple gradient detector of a local window located at a specified location $(x, y)$ of size $l \times l$ as: 
\begin{equation}
    \mathcal{D}_{xy}(I_{LR}) = \sum\nolimits_{i \in [x, x+l], j \in [y, y+l]}{\nabla_{ij}I_{LR}}.
\end{equation}

The path function $\gamma(\cdot)$ utilizes a Gaussian blur kernel to compute the blurred version of the input image, reducing the high-frequency components in the image to represent absent features:
\begin{equation}
    \gamma(\frac{k}{m}) = \omega(\sigma - \frac{k}{m} \sigma) \otimes I_{LR}
\end{equation}
where $\omega(\sigma)$ is the Gaussian kernel with the kernel width $\sigma$, and $\otimes$ denotes the convolution operation. Under this definition, $\gamma(\cdot)$ returns the original LR image $I_{LR}$ when $k=m$ and the completely blurred LR image $I'_{LR}$ when $k=0$, i.e., $\gamma(0) = I'_{LR}$ and $\gamma(1) = I_{LR}$.

The Diffusion Index (DI) can be derived from LAM as a quantitative indicator of pixel utilization. It is calculated based on the Gini coefficient $G$ measuring the inequality of pixels' impact:
\begin{gather}
 DI = (1 - G) \times 100 \\
 G = \frac{\sum_{i} \sum_{j} |g_i - g_j| }{2n^2\bar{g}}
\end{gather}
where $g_i$ represents the LAM value of pixel $i$, $n$ is the total number of pixels, and $\bar{g}$ is the average LAM value. Essentially, a high DI value indicates a model's capacity to involve a broader range of pixels in the generation of the target window, while a low DI value suggests a more limited involvement.

Although the LAM technique was initially developed for single image super-resolution, it can be seamlessly adapted for light field super-resolution without significant modifications because the BasicLFSR framework \cite{BasicLFSR} processes a 4D LF image as a 2D macro-pixel (MacPI) image \cite{wangDistgSSR_TIP2022}.
\end{highlight}

The LAM visualization and the DI are provided below PSNR/SSIM in Fig. \ref{fig:Qual}. The DI is calculated on the blue box regions with the red box regions as targets. The LAM results show that M2MT-Net consistently exhibits more activated pixels both within and across SAIs. This superiority is substantiated by the DI values as M2MT-Net is the highest, ranging between 19.7059 and 25.2688. This is 5-6\% higher than the second-ranked method, LF-DET, whose DI values range from 18.7618 to 23.9182. Meanwhile, the DI values of other competing methods are significantly lower, hovering from 5.1459 to 13.1903.

Delving deeper into these activated pixels reveals intriguing insights. For instance, in the \textit{Perforated\_Metal\_3} sample, though repeated perforated holes offer potential patterns for reconstruction, most methods focus solely on the neighboring area. LF-DET has some activated pixels on distance holes; however, the activation is weak. In contrast, M2MT-Net's activated pixels span not only the same column but also the neighboring columns with high activation, indicating that it identifies shared characteristics among the holes and leverages them as complementary cues. 
Similarly, in the \textit{Palais\_du\_Luxembourg} sample, the building's windows exhibit recurring patterns for reuse. M2MT-Net manages to utilize not only the windows in the red box but also the ones in a broader area of the blue box, and the influential pixels have high activities across SAIs. Hence, the patterns are recovered with visible edges, unlike its counterparts, which generate a blurry area due to their narrower focus and weaker correlation across SAIs.
For the \textit{Bicycle} sample, the plant leaves present similar patterns. M2MT-Net's advantage becomes evident as it activates pixels on leaves not only on the same trunk but also on the other trunk.

The DI values shed light on the relation between model performance and pixel utilization. In general, higher DI indicates higher pixel utilization and should result in better performance. It holds true for LF-DET and M2MT-Net as their DI values are significantly high as well as their PSNR and SSIM, and it remains consistent when comparing only within the convolution-based or Transformer-based groups. However, when comparing these two groups, a different trend emerges as a high DI does not necessarily mean high PSNR and SSIM, such as DistgSSR and LFT. This highlights the distinct nature of pixel utilization between convolutions and Transformers, where convolutions leverage more pixels but are constrained by locality, while Transformers establish long-range dependencies among broader pixels, though these dependencies may not always be strong enough to aid in super-resolution as effectively as M2MT-Net.

\begin{figure*}[ht!]
    \centering
    \tabcolsep=0.05cm
    \renewcommand{\arraystretch}{1.0}
    {\small 
    \begin{tabular}{cccccccc}
    \raisebox{-0.5\height}[0pt][0pt]{
        \includegraphics[width=0.20\textwidth, height=0.15\textwidth]{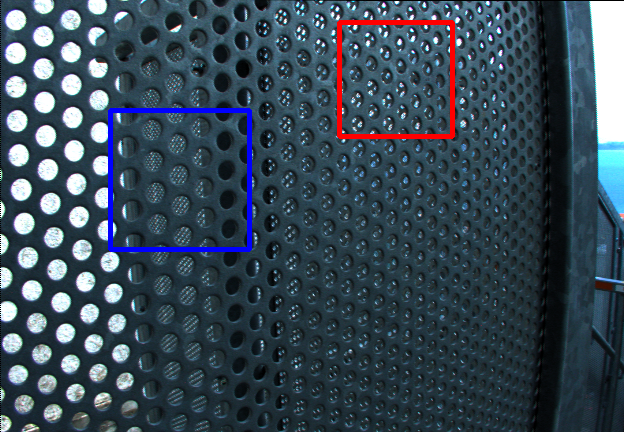}
    } &
    \includegraphics[width=0.100\textwidth, height=0.100\textwidth,cfbox=blue 1pt 0pt]{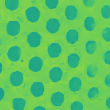} &
    \includegraphics[width=0.100\textwidth, height=0.100\textwidth,cfbox=blue 1pt 0pt]{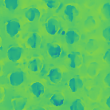} &
    \includegraphics[width=0.100\textwidth, height=0.100\textwidth,cfbox=blue 1pt 0pt]{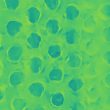} &
    \includegraphics[width=0.100\textwidth, height=0.100\textwidth,cfbox=blue 1pt 0pt]{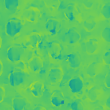} &
    \includegraphics[width=0.100\textwidth, height=0.100\textwidth,cfbox=blue 1pt 0pt]{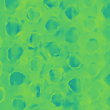} &
    \includegraphics[width=0.100\textwidth, height=0.100\textwidth,cfbox=blue 1pt 0pt]{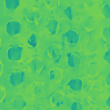} &
    \includegraphics[width=0.100\textwidth, height=0.100\textwidth,cfbox=blue 1pt 0pt]{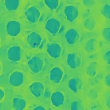} \\
    &
    \includegraphics[width=0.100\textwidth, height=0.100\textwidth,cfbox=red 1pt 0pt]{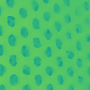} &
    \includegraphics[width=0.100\textwidth, height=0.100\textwidth,cfbox=red 1pt 0pt]{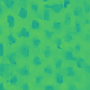} &
    \includegraphics[width=0.100\textwidth, height=0.100\textwidth,cfbox=red 1pt 0pt]{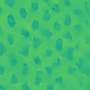} &
    \includegraphics[width=0.100\textwidth, height=0.100\textwidth,cfbox=red 1pt 0pt]{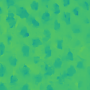} &
    \includegraphics[width=0.100\textwidth, height=0.100\textwidth,cfbox=red 1pt 0pt]{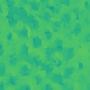} &
    \includegraphics[width=0.100\textwidth, height=0.100\textwidth,cfbox=red 1pt 0pt]{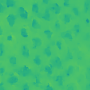} &
    \includegraphics[width=0.100\textwidth, height=0.100\textwidth,cfbox=red 1pt 0pt]{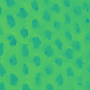} \\
    
    \textit{Perforated\_Metal\_3} &
    Ground-truth &
    DistgSSR &
    HLFSR &
    LFT &
    EPIT &
    LF-DET &
    M2MT-Net \\

    MSE$\times100$ &
    &
    0.831 &
    \underline{0.796} &
    1.046 &
    0.919 &
    0.834 &
    \textbf{0.613} \\

    \vspace{-5pt} \\

    \raisebox{-0.5\height}[0pt][0pt]{
        \includegraphics[width=0.20\textwidth, height=0.15\textwidth]{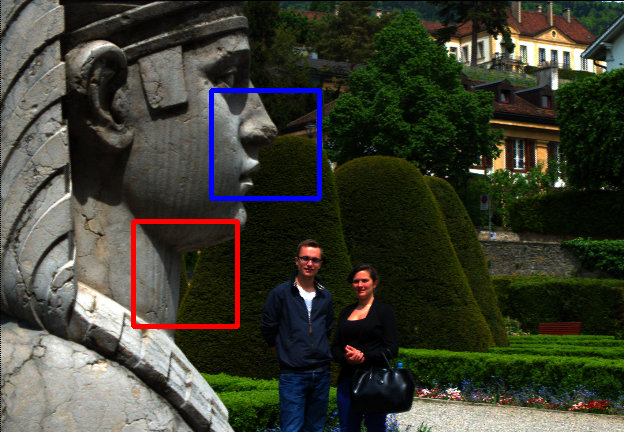}
    } &
    \includegraphics[width=0.100\textwidth, height=0.100\textwidth,cfbox=blue 1pt 0pt]{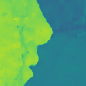} &
    \includegraphics[width=0.100\textwidth, height=0.100\textwidth,cfbox=blue 1pt 0pt]{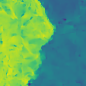} &
    \includegraphics[width=0.100\textwidth, height=0.100\textwidth,cfbox=blue 1pt 0pt]{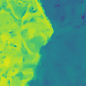} &
    \includegraphics[width=0.100\textwidth, height=0.100\textwidth,cfbox=blue 1pt 0pt]{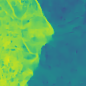} &
    \includegraphics[width=0.100\textwidth, height=0.100\textwidth,cfbox=blue 1pt 0pt]{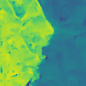} &
    \includegraphics[width=0.100\textwidth, height=0.100\textwidth,cfbox=blue 1pt 0pt]{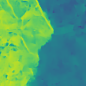} &
    \includegraphics[width=0.100\textwidth, height=0.100\textwidth,cfbox=blue 1pt 0pt]{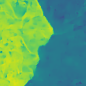} \\
    &
    \includegraphics[width=0.100\textwidth, height=0.100\textwidth,cfbox=red 1pt 0pt]{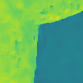} &
    \includegraphics[width=0.100\textwidth, height=0.100\textwidth,cfbox=red 1pt 0pt]{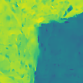} &
    \includegraphics[width=0.100\textwidth, height=0.100\textwidth,cfbox=red 1pt 0pt]{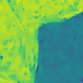} &
    \includegraphics[width=0.100\textwidth, height=0.100\textwidth,cfbox=red 1pt 0pt]{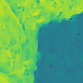} &
    \includegraphics[width=0.100\textwidth, height=0.100\textwidth,cfbox=red 1pt 0pt]{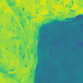} &
    \includegraphics[width=0.100\textwidth, height=0.100\textwidth,cfbox=red 1pt 0pt]{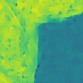} &
    \includegraphics[width=0.100\textwidth, height=0.100\textwidth,cfbox=red 1pt 0pt]{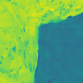} \\
    
    \textit{Sphynx} &
    Ground-truth &
    DistgSSR &
    HLFSR &
    LFT &
    EPIT &
    LF-DET &
    M2MT-Net \\

    MSE$\times100$ &
    &
    \underline{0.136} &
    0.150 &
    0.149 &
    0.146 &
    0.165 &
    \textbf{0.111} \\

    \vspace{-5pt} \\

    \raisebox{-0.5\height}[0pt][0pt]{
        \includegraphics[width=0.20\textwidth, height=0.18\textwidth]{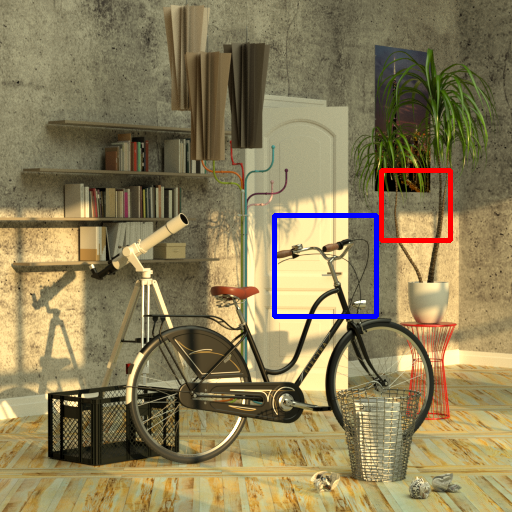}
    } &
    \includegraphics[width=0.100\textwidth, height=0.100\textwidth,cfbox=blue 1pt 0pt]{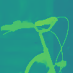} &
    \includegraphics[width=0.100\textwidth, height=0.100\textwidth,cfbox=blue 1pt 0pt]{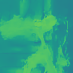} &
    \includegraphics[width=0.100\textwidth, height=0.100\textwidth,cfbox=blue 1pt 0pt]{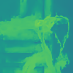} &
    \includegraphics[width=0.100\textwidth, height=0.100\textwidth,cfbox=blue 1pt 0pt]{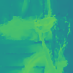} &
    \includegraphics[width=0.100\textwidth, height=0.100\textwidth,cfbox=blue 1pt 0pt]{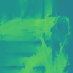} &
    \includegraphics[width=0.100\textwidth, height=0.100\textwidth,cfbox=blue 1pt 0pt]{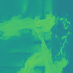} &
    \includegraphics[width=0.100\textwidth, height=0.100\textwidth,cfbox=blue 1pt 0pt]{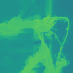} \\
    &
    \includegraphics[width=0.100\textwidth, height=0.100\textwidth,cfbox=red 1pt 0pt]{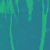} &
    \includegraphics[width=0.100\textwidth, height=0.100\textwidth,cfbox=red 1pt 0pt]{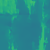} &
    \includegraphics[width=0.100\textwidth, height=0.100\textwidth,cfbox=red 1pt 0pt]{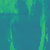} &
    \includegraphics[width=0.100\textwidth, height=0.100\textwidth,cfbox=red 1pt 0pt]{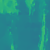} &
    \includegraphics[width=0.100\textwidth, height=0.100\textwidth,cfbox=red 1pt 0pt]{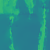} &
    \includegraphics[width=0.100\textwidth, height=0.100\textwidth,cfbox=red 1pt 0pt]{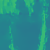} &
    \includegraphics[width=0.100\textwidth, height=0.100\textwidth,cfbox=red 1pt 0pt]{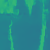} \\
    
    \textit{bicycle} &
    Ground-truth &
    DistgSSR &
    HLFSR &
    LFT &
    EPIT &
    LF-DET &
    M2MT-Net \\

    MSE$\times100$ &
    &
    2.627 &
    2.648 &
    2.653 &
    \underline{2.556} &
    2.609 &
    \textbf{2.224} \\

    \vspace{-5pt} \\

    \raisebox{-0.5\height}[0pt][0pt]{
        \includegraphics[width=0.20\textwidth, height=0.18\textwidth]{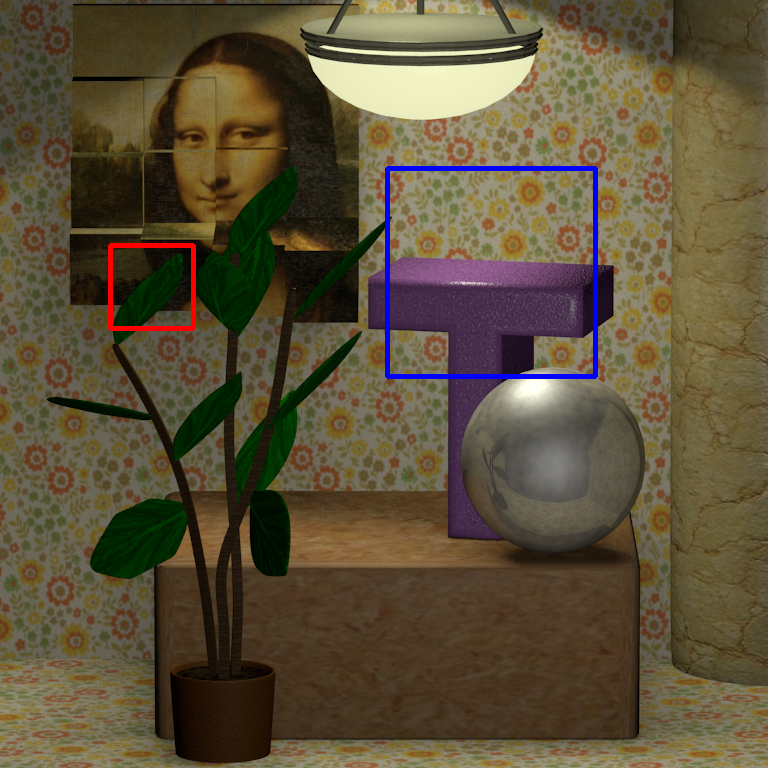}
    } &
    \includegraphics[width=0.100\textwidth, height=0.100\textwidth,cfbox=blue 1pt 0pt]{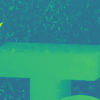} &
    \includegraphics[width=0.100\textwidth, height=0.100\textwidth,cfbox=blue 1pt 0pt]{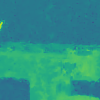} &
    \includegraphics[width=0.100\textwidth, height=0.100\textwidth,cfbox=blue 1pt 0pt]{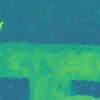} &
    \includegraphics[width=0.100\textwidth, height=0.100\textwidth,cfbox=blue 1pt 0pt]{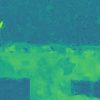} &
    \includegraphics[width=0.100\textwidth, height=0.100\textwidth,cfbox=blue 1pt 0pt]{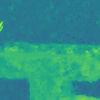} &
    \includegraphics[width=0.100\textwidth, height=0.100\textwidth,cfbox=blue 1pt 0pt]{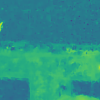} &
    \includegraphics[width=0.100\textwidth, height=0.100\textwidth,cfbox=blue 1pt 0pt]{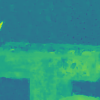} \\
    &
    \includegraphics[width=0.100\textwidth, height=0.100\textwidth,cfbox=red 1pt 0pt]{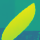} &
    \includegraphics[width=0.100\textwidth, height=0.100\textwidth,cfbox=red 1pt 0pt]{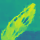} &
    \includegraphics[width=0.100\textwidth, height=0.100\textwidth,cfbox=red 1pt 0pt]{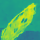} &
    \includegraphics[width=0.100\textwidth, height=0.100\textwidth,cfbox=red 1pt 0pt]{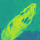} &
    \includegraphics[width=0.100\textwidth, height=0.100\textwidth,cfbox=red 1pt 0pt]{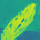} &
    \includegraphics[width=0.100\textwidth, height=0.100\textwidth,cfbox=red 1pt 0pt]{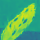} &
    \includegraphics[width=0.100\textwidth, height=0.100\textwidth,cfbox=red 1pt 0pt]{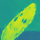} \\
    
    \textit{monasRoom} &
    Ground-truth &
    DistgSSR &
    HLFSR &
    LFT &
    EPIT &
    LF-DET &
    M2MT-Net \\

    MSE$\times100$ &
    &
    0.571 &
    \underline{0.556} &
    0.590 &
    0.567 &
    0.598 &
    \textbf{0.502} \\

    \end{tabular}
    }
    \caption{Visualization of depth estimation on the $4\times$ LFSR results of our M2MT-Net and the current state-of-the-art methods. Zoom-in depth maps are depicted on the areas in the blue and red boxes. MSE$\times 100$ is evaluated on the entire depth map. The best and second-best MSE are in bold and underlined.}
    \label{fig:Depth}
\end{figure*}

\subsection{Angular Consistency}
While the reconstruction of visual detail is important for LFSR, the preservation of parallax structure within LF images is equally crucial. This aspect cannot be adequately discerned solely by examining the reconstructed SAIs. Thus, to comprehensively assess the angular consistency, we conduct an evaluation through depth estimation. OACC-Net \cite{wangOACCNet_CVPR2022} is applied to generate depth maps on the super-resolved output of the methods under comparison. The depth estimated from HR images serves as the ground-truth for this evaluation. Fig. \ref{fig:Depth} visually represents the depth maps for two real-world and synthetic examples, accompanied by the $MSE\times100$ as a quantitative evaluation metric.

M2MT-Net's superiority, as highlighted in \textit{Perforated\_Metal\_3} of Fig. \ref{fig:Qual}, is corroborated in the generated depth map. This method successfully reconstructs more perforated holes with integrated edges spanning from near to distant from the camera as evidenced in the blue and red boxes. In stark contrast, competing methods struggle, yielding blurred and entangled edges in this complex scene. When examining scenes featuring salient objects, M2MT-Net continues to excel. In the \textit{Sphynx} sample, the contours of the sphynx's nose, mouth and neck are distinctly delineated in M2MT-Net's depth map. Other methods, however, generate noticeable blurs or artifacts in these areas. The \textit{bicycle} sample further illustrates M2MT-Net's proficiency, where distinct separations between the bicycle's handlebar and the background are evident, as well as more continuous structures of the plant's trunks. Other methods falter, blending the handlebar's contour with the background or breaking the trunk's structure into fragments. Finally, in the \textit{monasRoom} example, M2MT-Net's depth map reveals a smoother surface on the T-shaped object and integrated shapes of the leaf with fewer holes, demonstrating a closer approximation to the ground-truth when compared to the other methods, which produce noticeable bumpy artifacts.

These results collectively underscore M2MT-Net's leading capability not only in reconstructing visual details but also in preserving the parallax structure in super-resolved LF images, marking it as a significant advancement in LFSR.

\begin{table}[t!]
    \centering
    \begin{highlight}
    \tabcolsep=0.12cm
    \caption{Comparison of model efficiency and performance with the state-of-the-art methods by varying the number of blocks at the $4 \times$ scale. Time is the inference time. \#Params. is the number of parameters. Memory is the peak GPU memory usage for training. FLOPs is the number of floating-point operations. Time is the inference time. \#blocks is the number of blocks.}
    \label{tab:efficiency}
    
    \begin{tabular}{|l|c|c|c|c|c|c|}
    \hline
    Method                                                          & \makecell{\#Params.\\(M)}     & \makecell{Memory\\(GB)}      & \makecell{FLOPs\\(G)}     & \makecell{Time\\(s)}      & PSNR/SSIM      \\\hline\hline
    LF-IINet \cite{liuLFIINet_TMM2021}                              & 4.886                         & 1.99                         & 57.36                     & 1.55                      & 29.04/0.9188   \\
    LFSSR-SAV \cite{chengLFSSRSAV_TCI2022}                          & 1.542                         & 8.25                         & 99.45                     & 3.20                      & 29.37/0.9223   \\
    DistgSSR \cite{wangDistgSSR_TIP2022}                            & 3.582                         & 4.43                         & 65.26                     & 1.89                      & 28.99/0.9195   \\
    HLFSR \cite{duongHLFSR_TCI2023}                                 & 13.865                        & 2.43                         & 45.73                     & 6.83                      & 29.20/0.9222   \\\hline\hline
    \multicolumn{6}{|l|}{LFT \cite{liangLFT_SPL2022}} \\\hline
    \#blocks = 4                                                    & 1.163                         & 6.41                         & 30.20                     & 6.22                      & 29.33/0.9196   \\
    \#blocks = 8                                                    & 2.150                         & 11.11                        & 55.64                     & 12.27                     & 29.44/0.9219   \\
    \#blocks = 12                                                   & 3.136                         & 16.80                        & 81.07                     & 16.80                     & 29.59/0.9238   \\\hline\hline
    \multicolumn{6}{|l|}{EPIT \cite{liangEPIT_ICCV2023}} \\\hline
    \#blocks = 4                                                    & 1.212                         & 7.25                         & 57.87                     & 2.25                      & 29.20/0.9170   \\
    \#blocks = 5                                                    & 1.470                         & 8.63                         & 74.15                     & 2.61                      & 29.31/0.9196   \\
    \#blocks = 8                                                    & 2.246                         & 12.77                        & 110.98                    & 3.77                      & 29.50/0.9212   \\
    \#blocks = 12                                                   & 3.328                         & 18.30                        & 164.09                    & 5.28                      & 29.58/0.9212   \\
    \#blocks = 14                                                   & 3.797                         & 21.06                        & 190.65                    & 6.12                      & 29.53/0.9216   \\\hline\hline
    \multicolumn{6}{|l|}{LF-DET \cite{congLFDET_TMM2024}} \\\hline
    \#blocks = 3                                                    & 1.293                         & 13.80                        & 39.17                     & 3.79                      & 29.21/0.9199   \\
    \#blocks = 4                                                    & 1.697                         & 17.83                        & 51.20                     & 4.81                      & 29.42/0.9220   \\
    \#blocks = 5                                                    & 2.080                         & 21.86                        & 63.23                     & 5.91                      & 29.47/0.9228   \\\hline\hline
    \multicolumn{6}{|l|}{M2MT-Net (Ours)} \\\hline
    \#blocks = 3                                                    & 1.557                         & 2.72                         & 14.40                     & 1.14                      & 29.30/0.9222   \\
    \#blocks = 4                                                    & 2.043                         & 3.20                         & 18.29                     & 1.33                      & 29.50/0.9226   \\
    \#blocks = 5                                                    & 2.529                         & 3.67                         & 22.18                     & 1.52                      & 29.51/0.9239   \\
    \#blocks = 6                                                    & 3.015                         & 4.15                         & 26.07                     & 1.71                      & 29.58/0.9253   \\
    \#blocks = 7                                                    & 3.501                         & 4.62                         & 29.96                     & 1.90                      & 29.67/0.9265   \\
    \#blocks = 8                                                    & 3.986                         & 5.10                         & 33.85                     & 2.09                      & 29.85/0.9284   \\
    \#blocks = 9                                                    & 4.472                         & 5.57                         & 37.74                     & 2.28                      & 29.74/0.9259   \\\hline
    \end{tabular}
\end{highlight}
\end{table}

\begin{figure*}[ht]
    \centering
    \tabcolsep=0.01cm
    \scriptsize
    \renewcommand{\arraystretch}{1.5}
    \begin{tabular}{cccc}
        (a) PSNR vs Parameter Number &
        (b) PSNR vs Peak GPU Memory Usage &
        (c) PSNR vs FLOPs &
        (d) PSNR vs Inference Time \\

        \includegraphics[width=0.245\textwidth]{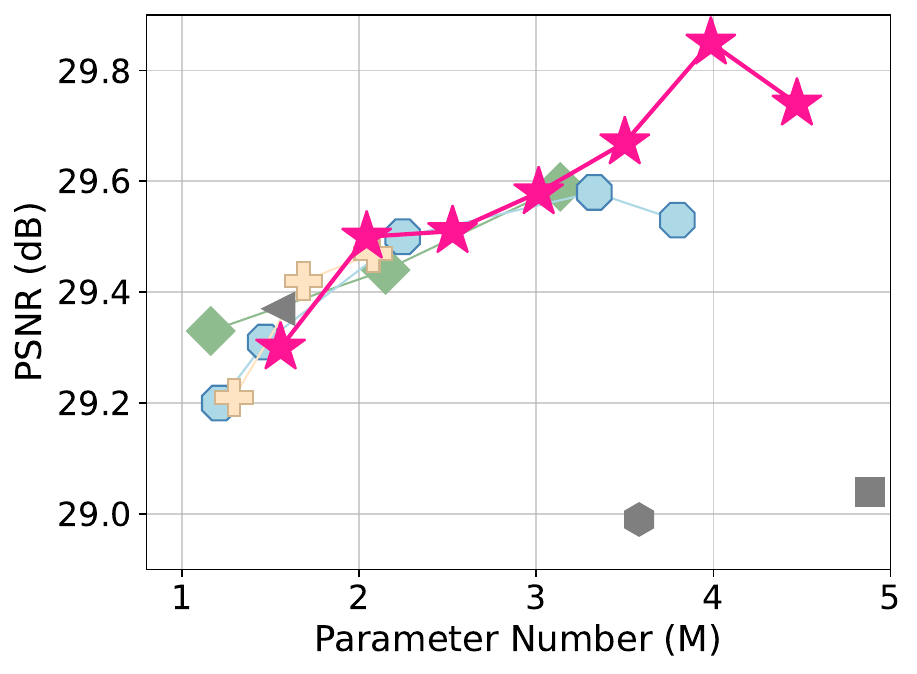} &
        \includegraphics[width=0.245\textwidth]{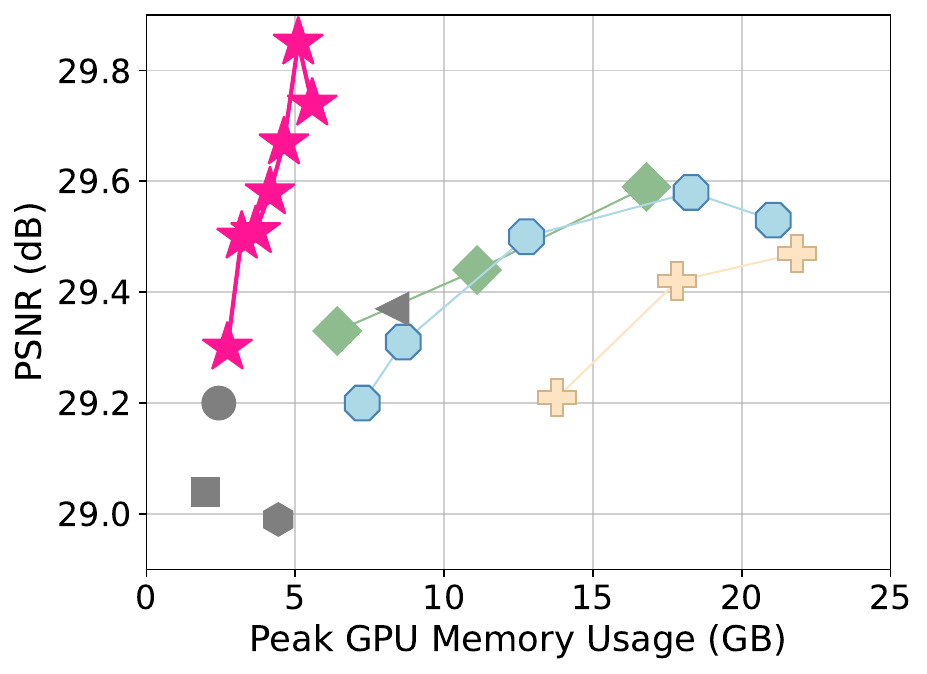} &
        \includegraphics[width=0.245\textwidth]{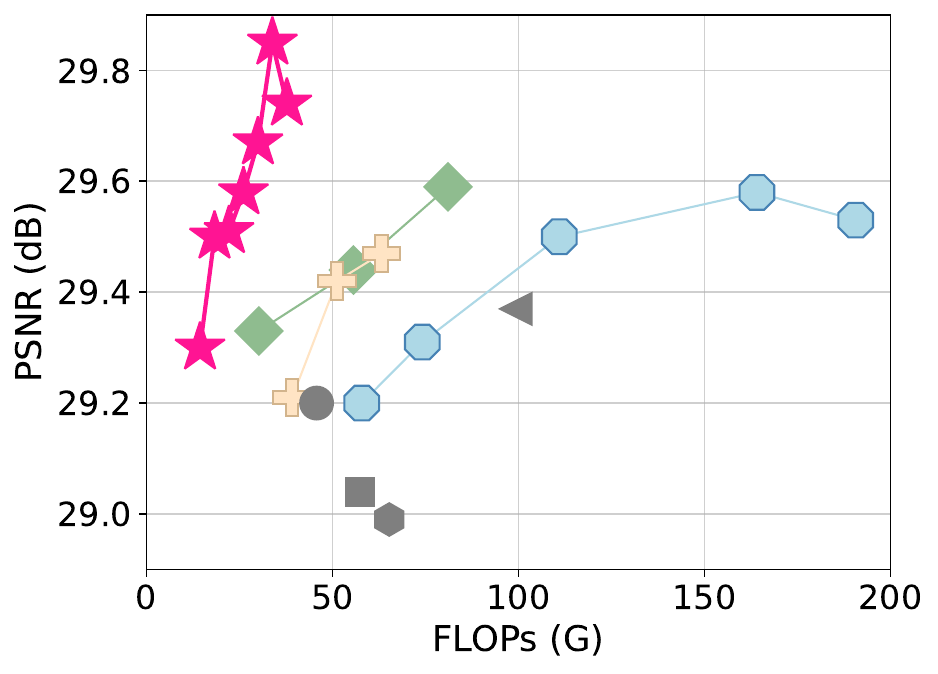} &
        \includegraphics[width=0.245\textwidth]{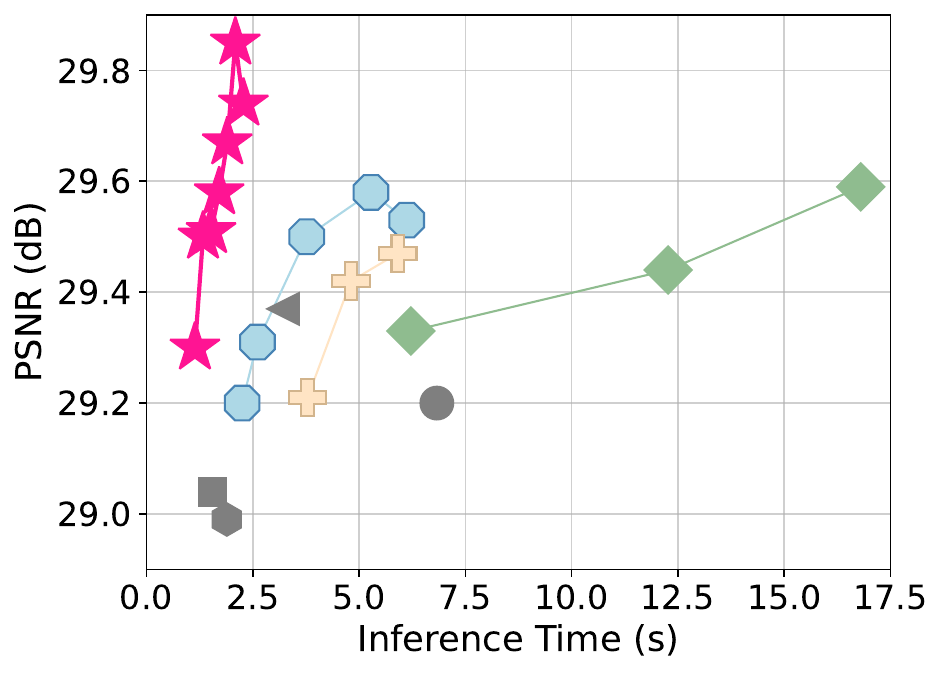} \\
        \multicolumn{4}{c}{
            \includegraphics[width=0.40\textwidth]{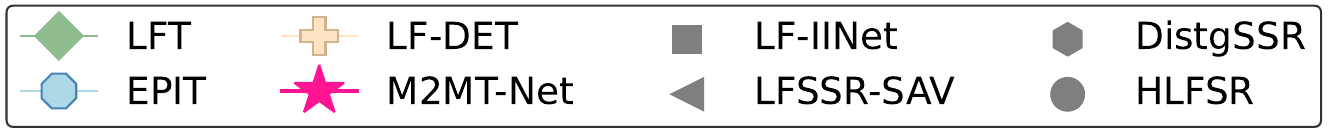}
        } 
    \end{tabular}
    \caption{\hl{Tradeoff between performance and efficiency at the $4\times$ scale. Candidates positioned closer to the top-left corner of the plots have a better performance-efficiency tradeoff. }}
    \label{fig:tradeoff}
\end{figure*}

\subsection{Model Efficiency}
\begin{highlight}
We evaluate the model efficiency of M2MT-Net against the top competitors, the convolution-based LF-IINet, LFSSR-SAV, DistgSSR and HLFSR, and the Transformer-based methods, LFT, EPIT and LF-DET at the $4 \times$ scale using four metrics: number of parameters, peak GPU memory usage, the floating-point operations (FLOPs), and inference time. The number of parameters and peak GPU memory reflect memory complexity, indicating the theoretical model size and the actual minimum memory required for training, respectively. FLOPs and inference time reflect computational complexity, representing the theoretical number of operations for processing a $32 \times 32$ LF patch and the actual average time for inferring a sample from the test datasets with GPU acceleration. FLOPs is obtained by utilizing the fvcore library \cite{fvcore}.

To ensure a fair and comprehensive evaluation, we train and evaluate variants of the Transformer-based methods, LFT, EPIT, LF-DET, and our M2MT-Net, varying the number of blocks (\#blocks, equivalent to $n_2$ in M2MT-Net). The results are compiled with PSNR on the \textit{EPFL} dataset in Table \ref{tab:efficiency}. To better understand how these methods balance performance and efficiency, we also plot the four efficiency metrics on the x-axis against PSNR on the y-axis in Fig. \ref{fig:tradeoff}. In these plots, models closer to the top-left corner represent a more favorable balance between performance and efficiency.

Regarding memory complexity, M2MT-Net is very similar to its Transformer-based peers in terms of the parameter number with minor PSNR differences under 0.1 dB among variants with similar parameter numbers in Fig. \ref{fig:tradeoff} (a). However, a stark contrast emerges in peak GPU memory usage as M2MT-Net's performance-efficiency curve (represented in pink) consistently trends toward the top-left direction relative to its competitors, LFT (in green), EPIT (in blue), and LF-DET (in yellow), in Fig. \ref{fig:tradeoff} (b). Specifically, M2MT-Net variants require less than 5.6 GB of GPU memory, whereas LFT starts at 6.41 GB, EPIT at 7.25 GB, and LF-DET at a hefty 13.80 GB. Notably, the 5-block LF-DET variant, while achieving a similar PSNR to the 4-block M2MT-Net (29.47 dB vs. 29.50 dB), consumes a significant 21.86 GB GPU memory. This is over six times the GPU memory used by the 4-block M2MT-Net (3.20 GB) and nearly maxes out the 24 GB memory of an Nvidia GTX 3090 GPU. LF-DET's high memory demand is primarily due to its complex design, which incorporates two spatial Transformers and three angular Transformers per block, which results in a memory bottleneck, significantly limiting LF-DET's scalability compared to its competitors. In contrast, M2MT-Net maintains low memory usage due to its streamlined and compact design.

From a computational complexity standpoint, M2MT-Net exhibits an exceptional performance-efficiency tradeoff in terms of FLOPs and inference time as its performance-efficiency curve consistently positions itself in the top-left direction relative to other methods' curves in Fig. \ref{fig:tradeoff} (c) and (d). The 6-block M2MT-Net, with a PSNR of 29.58 dB, matches or surpasses the best-performing variants of the other models, such as the 12-block LFT and EPIT (29.59 dB and 29.58 dB, respectively) and the 5-block LF-DET (29.47 dB). Remarkably, it requires only up to 41.23\% of the FLOPs (26.07 G) and 32.39\% of the inference time (1.71s) compared to the most efficient variants of these competitors (63.23 G by the 5-block LF-DET and 5.28s by the 12-block EPIT). Our top-performing 8-block M2MT-Net exceeds other methods by more than 0.26 dB in PSNR while still demanding lower FLOPs and inference time (33.85 G and 2.09s) than the most lightweight variants of nearly all other Transformer-based methods. On the other hand, while the PSNR of LFT demonstrates an upward trend with the addition of more blocks, its scalability is severely constrained due to its excessive inference time. Specifically, the 12-block variant of LFT requires an unmanageable inference time of 16.80 seconds, more than double that of its slowest competitor, the 14-block EPIT. This prohibitive inference duration renders further scaling of LFT impractical.

Meanwhile, the convolution-based methods generally exhibit a weaker performance-efficiency tradeoff as they are positioned lower and to the right compared to the Transformer-based methods, except LFSSR-SAV has a position similar to EPIT and LFT on the graph. Due to the inherent requirements of convolutional kernels, many convolution-based methods such as LF-IINet, DistgSSR, and HLFSR necessitate a larger number of parameters. This trend is illustrated in Fig. \ref{fig:tradeoff} (a). Notably, HLFSR requires more than 13.865 million parameters, which exceeds the graph's range.

In summary, these results demonstrate that M2MT-Net achieves excellent LFSR performance-efficiency balance and model scalability, making it a highly favorable choice for practical LFSR applications.
\end{highlight}

\subsection{Ablation Study}
\begin{table}[t]
    \centering
    \caption{Ablation study on altering components in correlation blocks. The best and second-best PSNR/SSIM are in bold and underlined.}
    \label{tab:ablation_altering}
    \resizebox{.45\textwidth}{!}{
    \begin{tabular}{|c|c|c|}
    \hline

    Spatial Component   &   Angular Component   & PSNR/SSIM \\ \hline
    M2MT                &   Vanilla Transformer & \textbf{29.85}/\textbf{0.9284} \\
    Vanilla Transformer &   Vanilla Transformer & 29.29/\underline{0.9213} \\
    Convolution         &   Vanilla Transformer & 29.02/0.9199 \\
    M2MT                &   Convolution         & \underline{29.42}/0.9208 \\
    \hline
    \end{tabular}
    }
\end{table}

In this section, we undertake a few ablation studies to understand the characteristics of M2MT-Net and its individual components. Note that the studies are carried out at the most challenging $4 \times$ scale using the \textit{EPFL} dataset with the most samples.

\subsubsection{Spatial and Angular Components} \label{section:ablation_altering}
To evaluate our proposed M2MT's role in the spatial subspace, we substitute it with a vanilla Transformer or a convolution and train the network. The modified networks have similar sizes to the original M2MT-Net to ensure a fair comparison. As indicated in Table \ref{tab:ablation_altering}, substituting the M2MT with a vanilla Transformer deteriorates the performance by 0.56 dB to 29.29 dB. Opting for a convolution results in a further decline, with a drop of 0.83 dB to 29.02 dB. These outcomes affirm M2MT's efficacy as a feature extractor in the spatial subspace compared to other alternatives when paired with its angular subspace counterpart.

We also train a M2MT-Net variant with the angular Transformer replaced with a convolution. Surprisingly, this variant achieves 29.42 dB PSNR, which is comparable to the best competitor, LF-DET, and outperforms other Transformer-based competitors like LFT and EPIT. This underscores the robust adaptability of the M2MT, even when paired with less potent components in the angular subspace.

\begin{highlight}
\subsubsection{Correlation Tensor Channels}
\begin{table}[t]
\centering
\begin{highlight}
    \tabcolsep=0.12cm
    \caption{Ablation study of M2MT-Net's correlation tensor channel number $C_{Cor}$. The best PSNR/SSIM are in bold.}
    \label{tab:correlation dimension}
    \begin{tabular}{|l|c|c|c|c|c|c|}
    \hline
    Method                                              & \makecell{FLOPs\\(G)}     & \makecell{Time\\(s)}      & \makecell{\#Params.\\(M)}     & \makecell{Memory\\(GB)}      & PSNR/SSIM                              \\\hline
    $C_{Cor} = 64$                                      & 31.21                     & 2.04                      & 2.460                         & 5.04                         & 29.60/0.9256                           \\
    $C_{Cor} = 96$                                      & 32.50                     & 2.05                      & 3.198                         & 5.08                         & 29.67/\underline{0.9265}               \\
    $C_{Cor} = 128$                                     & 33.85                     & 2.09                      & 3.986                         & 5.10                         & \textbf{29.85/0.9284}                  \\
    $C_{Cor} = 160$                                     & 35.24                     & 2.11                      & 4.824                         & 5.12                         & \underline{29.72}/\underline{0.9265}   \\\hline
    \end{tabular}
\end{highlight}
\end{table}
We evaluate the impact of varying the correlation tensor channel number $C_{Cor}$ on the performance and efficiency of M2MT-Net. Variants with $C_{Cor} \in \{64, 96, 160\}$ are trained and compared to the baseline $C_{Cor} = 128$, with results detailed in Table \ref{tab:correlation dimension}.

The PSNR results indicate an increase with the increment of $C_{Cor}$ from 64 to 128, starting from 29.60 dB and peaking at 29.85 dB. However, further increasing $C_{Cor}$ to 160 leads to a slight decrease in PSNR to 29.72 dB, which may be attributed to overfitting and redundancy in the correlation tensor. This suggests that the optimal value for $C_{Cor}$ is 128.

Additionally, the impact of $C_{Cor}$ on the model's memory and computational complexity is manageable. Increasing $C_{Cor}$ by 2.5 times from 64 to 160 results in a proportional 96.10\% increase in the parameter number, yet the increments in FLOPs, inference time, and peak GPU memory usage are relatively modest at 12.91\%, 3.43\%, and 1.59\%, respectively. These findings indicate that the correlation encoding and decoding processes are unlikely to be bottlenecks in the model's efficiency.
\end{highlight}

\section{Conclusion and Future Work}
In this paper, we have revealed the prevalent challenge of subspace isolation caused by the One-to-One scheme and present the novel concept of Many-to-Many Transformers (M2MT) as a new scheme to address this issue. The proposed M2MT is empowered with complete access to all pixels across all SAIs in a LF image to capture comprehensive long-range correlation dependencies. With M2MT as a pivotal component, we have proposed a simple yet effective M2MT-Net for LFSR. Extensive experiments on various public datasets have demonstrated that M2MT-Net surpasses state-of-the-art methods in terms of reconstructed image quality while maintaining favorable computational and memory efficiency, making it a viable model for practical LF applications. Further, our analysis of angular consistency through LF depth estimation shows that M2MT-Net not only reconstructs finer visual details but also preserves and enhances the parallax structure of LF images. Its superiority is evidenced by visual interpretability in our in-depth analysis using the LAM technique, which highlights that M2MT involves a substantially broader range of pixels across wider SAIs beyond subspace isolation, signifying its truly global context and a more comprehensive modeling of correlation dependencies.

\begin{highlight}
Looking ahead, there are some promising directions for improving M2MT in future works:
\begin{enumerate}
    \item \textbf{More Subspaces.} Enhancing M2MT's capacity by extending it to subspaces like the EPI can address large disparities, as seen in datasets like \textit{STFgantry} \cite{vaishSTFgantry_2008}, akin to EPIT \cite{liangEPIT_ICCV2023}. Additionally, applying M2MT to the angular subspace could create a symmetric structure with its existing spatial counterpart. Nonetheless, two main challenges are anticipated: Firstly, the unique characteristics of these subspaces may require specific modifications to the Many-to-Many mechanism. Secondly, increasing model complexity becomes a significant concern when either spatial dimension, $W$ or $H$, is engaged as embeddings in the correlation encoding and decoding processes. Currently, these processes are achieved by manageable linear layers between $UVC$ and $C_{Cor}$. As $W \gg U$ and $H \gg V$, directly replacing $U$ or $V$ with $W$ or $H$ leads to unmanageable memory and computational complexities.
    \item \textbf{Light Field View Synthesis.} M2MT holds potential for application in LF view synthesis. Particularly, some LF view synthesis methods \cite{wuSAAN_TIP2021, wuShearedEPI_TIP2019,chengLFSSRSAV_TCI2022} employ EPI-based strategies by extracting correlation features from EPIs or super-resolving EPIs to super-resolve the angular resolution. As discussed in the first point, enabling M2MT to operate within the EPI subspaces could effectively leverage its capabilities for this task. 
    \item \textbf{Unified M2MT.} It will be a compelling advancement to unify the spatial and angular Transformers into a single and holistic M2MT component. This integration would enable simultaneous and cohesive execution of spatial and angular self-attention processes, likely improving the model's efficiency and effectiveness with compact spatial-angular feature extraction.
\end{enumerate}
\end{highlight}

\bibliographystyle{IEEEtran}
\bibliography{reference}

\begin{IEEEbiography}[{\includegraphics[width=1in,height=1.25in,clip,keepaspectratio]{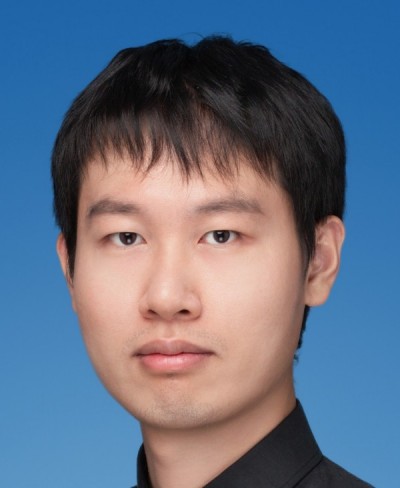}}]{Zeke Zexi Hu} is currently a Ph.D. candidate at the School of Computer Science, University of Sydney, Australia. He received his bachelor's degree from South China Agricultural University, China in 2014 and his M.Phil. degree from the University of Sydney in 2020. His research focuses on computer vision and deep learning. He has authored and co-authored papers in academic journals and conferences, including TCSVT, TVCG, TIM, ICIP, etc.
\end{IEEEbiography}

\begin{IEEEbiography}[{\includegraphics[width=1in,height=1.25in,clip,keepaspectratio]{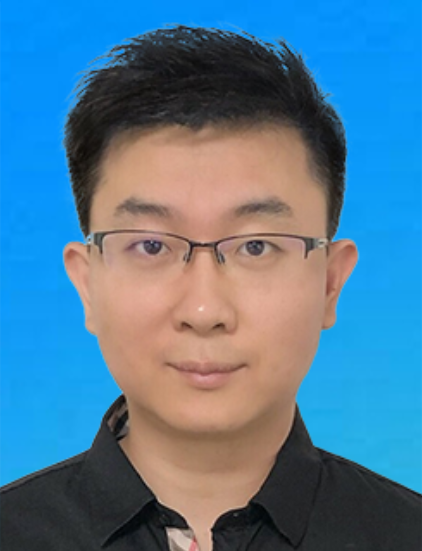}}]{Xiaoming Chen} holds a B.Sc. degree (with Distinction) from Royal Melbourne Institute of Technology and a Ph.D. degree from the University of Sydney, Australia. He has a combined experience in industry and academia. He has been with National University of Singapore, Nanyang Technological University, Singapore, CSIRO Australia, Technicolor Research, IBM Corporation, and University of Science and Technology of China (USTC). He is now a Professor at the School of Computer and Artificial Intelligence, Beijing Technology and Business University (BTBU), China, and a researcher at the University of Sydney, Australia. His research interests include immersive media computing, virtual reality, bio-inspired event processing, and related applications. His work has been published in journals and conferences including IEEE Trans. Vis. Comput. Graph., IEEE Trans. Image Process., IEEE Trans. Circuits Syst. Video Technol., IEEE Trans. Mult., IEEE VR, ACM MM, ECCV, AAAI, etc.
\end{IEEEbiography}

\begin{IEEEbiography}[{\includegraphics[width=1in,height=1.25in,clip,keepaspectratio]{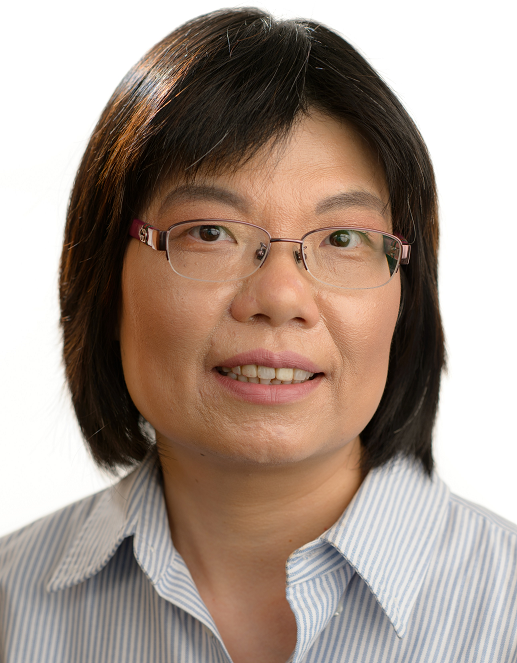}}]{Vera Yuk Ying Chung} received her B.S. degree in Computing and Information Systems from the University of London, UK, in 1995, and her Ph.D. degree in Computer Engineering from the Queensland University of Technology, Australia in 2000. She is a Senior Lecturer at the School of Computer Science, University of Sydney, Australia. Her research interests are in the areas of multimedia computing and virtual reality. Her work has been published in journals and conferences including IEEE Trans. on Image Processing, IEEE Trans. Vis. Comput. Graph., IEEE Trans. Circuits Syst. Video Technol., NeurIPS, ECCV, etc.
\end{IEEEbiography}

\begin{IEEEbiography}[{\includegraphics[width=1in,height=1.25in,clip,keepaspectratio]{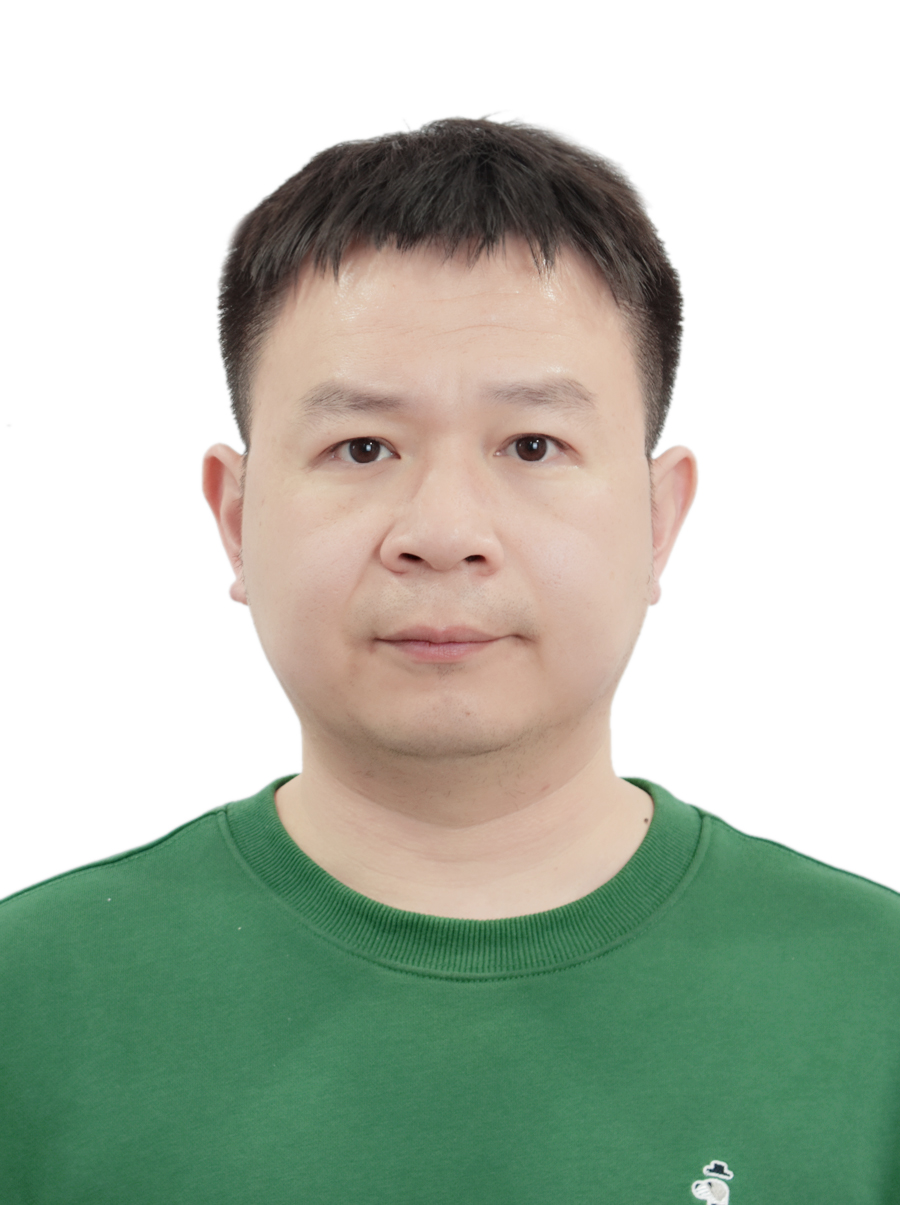}}]
{Yiran Shen} is professor in School of Software, Shandong University. He received his BE in communication engineering from Shandong University, China and his PhD degree in computer science and engineering from University of New South Wales. He published regularly at top-tier conferences and journals. Generally speaking, his research interest is sensing and computing for immersive systems. He is a senior member of IEEE.
\end{IEEEbiography}

\end{document}